%% file: main-tyde.tex
\documentclass[sigplan]{acmart}
\setcopyright{licensedothergov}
\acmPrice{15.00}
\acmDOI{10.1145/3122975.3122978}
\acmYear{2017}
\copyrightyear{2017}
\acmISBN{978-1-4503-5183-6/17/09}
\acmConference[TyDe'17]{2nd ACM SIGPLAN International Workshop on Type-Driven Development}{September 3, 2017}{Oxford, UK}

\usepackage{amsmath}
\usepackage{xcolor}
\usepackage{url}
\usepackage{doubleequals}
\usepackage{textcomp} 
\usepackage{fancyvrb} 
\usepackage{flushend} 

\input{Preamble}
\input{macro}

\begin{document}

\title{Type Safe Redis Queries:\\A Case Study of Type-Level Programming in Haskell}


\author{Ting-Yan Lai}
\authornote{Ting-Yan Lai was a summer intern from
the Inst. of Computer Science and Engineering,
National Chiao Tung Univ., Taiwan,
when part of this work was developed.}
\affiliation{
  \department{Institute of Information Science}
  \institution{Academia Sinica}
  \streetaddress{128, Academia Road, Section 2}
  \city{Nangang, Taipei City}
  \postcode{115}
  \country{Taiwan}
}
\email{banacorn@gmail.com}          

\author{Tyng-Ruey Chuang}
\affiliation{
  \department{Institute of Information Science}
  \institution{Academia Sinica}
  \streetaddress{128, Academia Road, Section 2}
  \city{Nangang, Taipei City}
  \postcode{115}
  \country{Taiwan}
}
\email{trc@iis.sinica.edu.tw}

\author{Shin-Cheng Mu}
\affiliation{
  \department{Institute of Information Science}
  \institution{Academia Sinica}
  \streetaddress{128, Academia Road, Section 2}
  \city{Nangang, Taipei City}
  \postcode{115}
  \country{Taiwan}
}
\email{scm@iis.sinica.edu.tw}

\begin{abstract}
\Redis{} is an in-memory data structure store, often used as a database, with a
Haskell interface \Hedis{}. \Redis{} is dynamically typed --- a key can be
discarded and re-associated to a value of a different type, and a command,
when fetching a value of a type it does not expect, signals a runtime error. We
develop a domain-specific language that, by exploiting Haskell type-level
programming techniques including indexed monad, type-level literals and closed
type families, keeps track of types of values in the database and statically
guarantees that type errors cannot happen for a class of \Redis{} programs.
\end{abstract}

\ccsdesc[500]{Software and its engineering~Functional languages}
\ccsdesc[300]{Software and its engineering~Domain specific languages}
\ccsdesc[300]{Information systems~Query languages}

\keywords{Haskell, Redis, Type-Level Programming, Type Safety, Key-Value Store}

\maketitle
\thispagestyle{empty}

\input{sections/Introduction}
\input{sections/IndexedMonads}
\input{sections/TypeLevelDict}
\input{sections/Embedding}
\input{sections/Discussions}

\input{sections/Conclusions}

\bibliography{cites}

\end{document}

%% file: macro.tex
\newcommand{\hide}[1]{}
\newcommand{\todo}[1]{\noindent\textcolor{blue}{\ifmmode \text{[#1]}\else [#1] \fi}}

\newcommand{\Redis}{{\sc Redis}}
\newcommand{\Hedis}{{\sc Hedis}}
\newcommand{\Edis}{{\sc Edis}}

%% file: sections/TypeLevelDict.tex
%
%
\makeatletter
\@ifundefined{lhs2tex.lhs2tex.sty.read}%
  {\@namedef{lhs2tex.lhs2tex.sty.read}{}%
   \newcommand\SkipToFmtEnd{}%
   \newcommand\EndFmtInput{}%
   \long\def\SkipToFmtEnd#1\EndFmtInput{}%
  }\SkipToFmtEnd

\newcommand\ReadOnlyOnce[1]{\@ifundefined{#1}{\@namedef{#1}{}}\SkipToFmtEnd}
\usepackage{amstext}
\usepackage{amssymb}
\usepackage{stmaryrd}
\DeclareFontFamily{OT1}{cmtex}{}
\DeclareFontShape{OT1}{cmtex}{m}{n}
  {<5><6><7><8>cmtex8
   <9>cmtex9
   <10><10.95><12><14.4><17.28><20.74><24.88>cmtex10}{}
\DeclareFontShape{OT1}{cmtex}{m}{it}
  {<-> ssub * cmtt/m/it}{}
\newcommand{\texfamily}{\fontfamily{cmtex}\selectfont}
\DeclareFontShape{OT1}{cmtt}{bx}{n}
  {<5><6><7><8>cmtt8
   <9>cmbtt9
   <10><10.95><12><14.4><17.28><20.74><24.88>cmbtt10}{}
\DeclareFontShape{OT1}{cmtex}{bx}{n}
  {<-> ssub * cmtt/bx/n}{}
\newcommand{\tex}[1]{\text{\texfamily#1}}	

\newcommand{\Sp}{\hskip.33334em\relax}

\newcommand{\Conid}[1]{\mathit{#1}}
\newcommand{\Varid}[1]{\mathit{#1}}
\newcommand{\anonymous}{\kern0.06em \vbox{\hrule\@width.5em}}
\newcommand{\plus}{\mathbin{+\!\!\!+}}
\newcommand{\bind}{\mathbin{>\!\!\!>\mkern-6.7mu=}}
\newcommand{\rbind}{\mathbin{=\mkern-6.7mu<\!\!\!<}}
\newcommand{\sequ}{\mathbin{>\!\!\!>}}
\renewcommand{\leq}{\leqslant}
\renewcommand{\geq}{\geqslant}
\usepackage{polytable}

\@ifundefined{mathindent}%
  {\newdimen\mathindent\mathindent\leftmargini}%
  {}%

\def\resethooks{%
  \global\let\SaveRestoreHook\empty
  \global\let\ColumnHook\empty}
\newcommand*{\savecolumns}[1][default]%
  {\g@addto@macro\SaveRestoreHook{\savecolumns[#1]}}
\newcommand*{\restorecolumns}[1][default]%
  {\g@addto@macro\SaveRestoreHook{\restorecolumns[#1]}}
\newcommand*{\aligncolumn}[2]%
  {\g@addto@macro\ColumnHook{\column{#1}{#2}}}

\resethooks

\newcommand{\onelinecommentchars}{\quad-{}- }
\newcommand{\commentbeginchars}{\enskip\{-}
\newcommand{\commentendchars}{-\}\enskip}

\newcommand{\visiblecomments}{%
  \let\onelinecomment=\onelinecommentchars
  \let\commentbegin=\commentbeginchars
  \let\commentend=\commentendchars}

\newcommand{\invisiblecomments}{%
  \let\onelinecomment=\empty
  \let\commentbegin=\empty
  \let\commentend=\empty}

\visiblecomments

\newlength{\blanklineskip}
\setlength{\blanklineskip}{0.66084ex}

\newcommand{\hsindent}[1]{\quad}
\let\hspre\empty
\let\hspost\empty
\newcommand{\NB}{\textbf{NB}}
\newcommand{\Todo}[1]{$\langle$\textbf{To do:}~#1$\rangle$}

\EndFmtInput
\makeatother
%
%
%
%
%
%
%
%
%
\ReadOnlyOnce{polycode.fmt}%
\makeatletter

\newcommand{\hsnewpar}[1]%
  {{\parskip=0pt\parindent=0pt\par\vskip #1\noindent}}

\newcommand{\hscodestyle}{}


\newcommand{\sethscode}[1]%
  {\expandafter\let\expandafter\hscode\csname #1\endcsname
   \expandafter\let\expandafter\endhscode\csname end#1\endcsname}


\newenvironment{compathscode}%
  {\par\noindent
   \advance\leftskip\mathindent
   \hscodestyle
   \let\\=\@normalcr
   \let\hspre\(\let\hspost\)%
   \pboxed}%
  {\endpboxed\)%
   \par\noindent
   \ignorespacesafterend}

\newcommand{\compaths}{\sethscode{compathscode}}


\newenvironment{plainhscode}%
  {\hsnewpar\abovedisplayskip
   \advance\leftskip\mathindent
   \hscodestyle
   \let\hspre\(\let\hspost\)%
   \pboxed}%
  {\endpboxed%
   \hsnewpar\belowdisplayskip
   \ignorespacesafterend}

\newenvironment{oldplainhscode}%
  {\hsnewpar\abovedisplayskip
   \advance\leftskip\mathindent
   \hscodestyle
   \let\\=\@normalcr
   \(\pboxed}%
  {\endpboxed\)%
   \hsnewpar\belowdisplayskip
   \ignorespacesafterend}


\newcommand{\plainhs}{\sethscode{plainhscode}}
\newcommand{\oldplainhs}{\sethscode{oldplainhscode}}
\plainhs


\newenvironment{arrayhscode}%
  {\hsnewpar\abovedisplayskip
   \advance\leftskip\mathindent
   \hscodestyle
   \let\\=\@normalcr
   \(\parray}%
  {\endparray\)%
   \hsnewpar\belowdisplayskip
   \ignorespacesafterend}

\newcommand{\arrayhs}{\sethscode{arrayhscode}}


\newenvironment{mathhscode}%
  {\parray}{\endparray}

\newcommand{\mathhs}{\sethscode{mathhscode}}


\newenvironment{texthscode}%
  {\(\parray}{\endparray\)}

\newcommand{\texths}{\sethscode{texthscode}}


\def\codeframewidth{\arrayrulewidth}
\RequirePackage{calc}

\newenvironment{framedhscode}%
  {\parskip=\abovedisplayskip\par\noindent
   \hscodestyle
   \arrayrulewidth=\codeframewidth
   \tabular{@{}|p{\linewidth-2\arraycolsep-2\arrayrulewidth-2pt}|@{}}%
   \hline\framedhslinecorrect\\{-1.5ex}%
   \let\endoflinesave=\\
   \let\\=\@normalcr
   \(\pboxed}%
  {\endpboxed\)%
   \framedhslinecorrect\endoflinesave{.5ex}\hline
   \endtabular
   \parskip=\belowdisplayskip\par\noindent
   \ignorespacesafterend}

\newcommand{\framedhslinecorrect}[2]%
  {#1[#2]}

\newcommand{\framedhs}{\sethscode{framedhscode}}


\newenvironment{inlinehscode}%
  {\(\def\column##1##2{}%
   \let\>\undefined\let\<\undefined\let\\\undefined
   \newcommand\>[1][]{}\newcommand\<[1][]{}\newcommand\\[1][]{}%
   \def\fromto##1##2##3{##3}%
   \def\nextline{}}{\) }%

\newcommand{\inlinehs}{\sethscode{inlinehscode}}


\newenvironment{joincode}%
  {\let\orighscode=\hscode
   \let\origendhscode=\endhscode
   \def\endhscode{\def\hscode{\endgroup\def\@currenvir{hscode}\\}\begingroup}
   \orighscode\def\hscode{\endgroup\def\@currenvir{hscode}}}%
  {\origendhscode
   \global\let\hscode=\orighscode
   \global\let\endhscode=\origendhscode}%

\makeatother
\EndFmtInput

\ReadOnlyOnce{Formatting.fmt}%
\makeatletter

\let\Varid\mathit
\let\Conid\mathsf

\def\commentbegin{\quad\{\ }
\def\commentend{\}}

\newcommand{\ty}[1]{\Conid{#1}}
\newcommand{\con}[1]{\Conid{#1}}
\newcommand{\id}[1]{\Varid{#1}}
\newcommand{\cl}[1]{\Varid{#1}}
\newcommand{\opsym}[1]{\mathrel{#1}}

\newcommand\Keyword[1]{\textbf{\textsf{#1}}}
\newcommand\Hide{\mathbin{\downarrow}}
\newcommand\Reveal{\mathbin{\uparrow}}


\makeatother
\EndFmtInput

\section{Type-Level Dictionaries}
\label{sec:type-level-dict}

One of the challenges of statically ensuring type correctness of stateful
languages is that the type of the value of a key can be altered by updating.
In \Redis{}, one may delete an existing key and create it again by assigning to
it a value of a different type. To ensure type correctness, we keep track of the
types of all existing keys in a {\em dictionary}. A dictionary is
a finite map, which can be represented by an associate list, or a
list of pairs of keys and some encoding of types. For example, we may use the
dictionary \ensuremath{[\mskip1.5mu (\text{\tt \char34 A\char34},\Conid{Int}),(\text{\tt \char34 B\char34},\Conid{Char}),(\text{\tt \char34 C\char34},\Conid{Bool})\mskip1.5mu]} to represent a
predicate, or a constraint, stating that ``the keys in the data store are \ensuremath{\text{\tt \char34 A\char34}},
\ensuremath{\text{\tt \char34 B\char34}}, and \ensuremath{\text{\tt \char34 C\char34}}, respectively assigned values of type \ensuremath{\Conid{Int}}, \ensuremath{\Conid{Char}}, and
\ensuremath{\Conid{Bool}}.'' (This representation will be refined in the next section.)

The dictionary above mixes values (strings such as \ensuremath{\text{\tt \char34 A\char34}}, \ensuremath{\text{\tt \char34 B\char34}}) and types.
Furthermore, as mentioned in Section~\ref{sec:indexed-monads}, the
dictionaries will be parameters to the indexed monad \ensuremath{\Conid{Edis}}. In a dependently
typed programming language (without the so-called ``phase distinction'' ---
separation between types and terms), this would pose no problem. In Haskell
however, the dictionaries, to index a monad, has to be a type as well.

In this section we describe how to construct a type-level dictionary, to be
used with the indexed monad in Section~\ref{sec:indexed-monads}.

\subsection{Datatype Promotion}

Haskell maintains the distinction between values, types, and kinds: values are
categorized by types, and types are categorized by kinds. The kinds are
relatively simple: \ensuremath{\mathbin{*}} is the kind of all {\em lifted} types, while type
constructors have kinds such as \ensuremath{\mathbin{*}\to \mathbin{*}}, or \ensuremath{\mathbin{*}\to \mathbin{*}\to \mathbin{*}}, etc.\footnotemark\,
Consider the datatype definitions below:
\begin{hscode}\SaveRestoreHook
\column{B}{@{}>{\hspre}l<{\hspost}@{}}%
\column{E}{@{}>{\hspre}l<{\hspost}@{}}%
\>[B]{}\mathbf{data}\;\Conid{Nat}\mathrel{=}\Conid{Zero}\mid \Conid{Suc}\;\Conid{Nat}~~,\qquad\;\mathbf{data}\;[\mskip1.5mu \Varid{a}\mskip1.5mu]\mathrel{=}[\mskip1.5mu \mskip1.5mu]\mid \Varid{a}\mathbin{:}[\mskip1.5mu \Varid{a}\mskip1.5mu]~~.{}\<[E]%
\ColumnHook
\end{hscode}\resethooks
The left-hand side is usually seen as having defined a type \ensuremath{\Conid{Nat}\mathbin{::}\mathbin{*}},
and two value constructors \ensuremath{\Conid{Zero}\mathbin{::}\Conid{Nat}} and \ensuremath{\Conid{Suc}\mathbin{::}\Conid{Nat}\to \Conid{Nat}}. The right-hand
side is how Haskell lists are understood. The {\em kind} of \ensuremath{[\mskip1.5mu \mathbin{\cdot}\mskip1.5mu]} is \ensuremath{\mathbin{*}\to \mathbin{*}},
since it takes a lifted type \ensuremath{\Varid{a}} to a lifted type \ensuremath{[\mskip1.5mu \Varid{a}\mskip1.5mu]}. The two value
constructors respectively have types \ensuremath{[\mskip1.5mu \mskip1.5mu]\mathbin{::}[\mskip1.5mu \Varid{a}\mskip1.5mu]} and \ensuremath{(\mathbin{:})\mathbin{::}\Varid{a}\to [\mskip1.5mu \Varid{a}\mskip1.5mu]\to [\mskip1.5mu \Varid{a}\mskip1.5mu]}, for all types \ensuremath{\Varid{a}}.

\footnotetext{In Haskell, the opposite of \emph{lifted} types are \emph{unboxed}
types, which are not represented by a pointer to a heap object, and cannot be
stored in a polymorphic data type.}

The GHC extension \emph{data kinds}~\cite{promotion}, however, automatically
promotes certain ``suitable'' types to kinds.\footnote{It is only informally
described in the GHC manual what types are ``suitable''.} With the extension,
the \ensuremath{\mathbf{data}} definitions above has an alternative reading: \ensuremath{\Conid{Nat}} is a new kind,
\ensuremath{\Conid{Zero}\mathbin{::}\Conid{Nat}} is a type having kind \ensuremath{\Conid{Nat}}, and \ensuremath{\Conid{Suc}\mathbin{::}\Conid{Nat}\to \Conid{Nat}} is a type
constructor, taking a type in kind \ensuremath{\Conid{Nat}} to another type in \ensuremath{\Conid{Nat}}. When one
sees a constructor in an expression, whether it is promoted can often be
inferred from the context. When one needs to be more specific, prefixing a
constructor with a single quote, such as in \ensuremath{\mbox{\textquotesingle}\Conid{Zero}} and \ensuremath{\mbox{\textquotesingle}\Conid{Suc}}, denotes that it
is promoted.

The situation of lists is similar: for all kinds \ensuremath{\Varid{k}}, \ensuremath{[\mskip1.5mu \Varid{k}\mskip1.5mu]} is also a kind. For
all kinds \ensuremath{\Varid{k}}, \ensuremath{\mbox{\textquotesingle}[~]} is a type of kind \ensuremath{[\mskip1.5mu \Varid{k}\mskip1.5mu]}. Given a type \ensuremath{\Varid{a}} of kind \ensuremath{\Varid{k}} and a
type \ensuremath{\Varid{as}} of kind \ensuremath{[\mskip1.5mu \Varid{k}\mskip1.5mu]}, \ensuremath{\Varid{a}\mathrel{\,\mbox{\textquotesingle}\!\!:}\Varid{as}} is again a type of kind \ensuremath{[\mskip1.5mu \Varid{k}\mskip1.5mu]}. Formally,
\ensuremath{(\mathrel{\,\mbox{\textquotesingle}\!\!:})\mathbin{::}\Varid{k}\to [\mskip1.5mu \Varid{k}\mskip1.5mu]\to [\mskip1.5mu \Varid{k}\mskip1.5mu]}. For example, \ensuremath{\Conid{Int}\mathrel{\,\mbox{\textquotesingle}\!\!:}(\Conid{Char}\mathrel{\,\mbox{\textquotesingle}\!\!:}(\Conid{Bool}\mathrel{\,\mbox{\textquotesingle}\!\!:}\mbox{\textquotesingle}[~]))} is a
type having kind \ensuremath{[\mskip1.5mu \mathbin{*}\mskip1.5mu]} --- it is a list of (lifted) types. The optional quote
denotes that the constructors are promoted. The same list can be denoted by a
syntax sugar \ensuremath{\mbox{\textquotesingle}[\Conid{Int},\Conid{Char},\Conid{Bool}]}.

Tuples are also promoted. Thus we may put two types in a pair to form another
type, such as in \ensuremath{\mbox{\textquotesingle}(\Conid{Int},\Conid{Char})}, a type having kind \ensuremath{(\mathbin{*},\mathbin{*})}.

Strings in Haskell are nothing but lists of \ensuremath{\Conid{Char}}s. Regarding promotion,
however, a string can be promoted to a type having kind \ensuremath{\Conid{Symbol}}. In the expression:
\begin{hscode}\SaveRestoreHook
\column{B}{@{}>{\hspre}l<{\hspost}@{}}%
\column{E}{@{}>{\hspre}l<{\hspost}@{}}%
\>[B]{}\text{\tt \char34 this~is~a~type-level~string~literal\char34}\mathbin{::}\Conid{Symbol}~~,{}\<[E]%
\ColumnHook
\end{hscode}\resethooks
the string on the left-hand side of \ensuremath{(\mathbin{::})} is a type whose kind is \ensuremath{\Conid{Symbol}}.

With all of these ingredients, we are ready to build our dictionaries, or
type-level associate lists:
\begin{hscode}\SaveRestoreHook
\column{B}{@{}>{\hspre}l<{\hspost}@{}}%
\column{E}{@{}>{\hspre}l<{\hspost}@{}}%
\>[B]{}\mathbf{type}\;\Conid{DictEmpty}\mathrel{=}\mbox{\textquotesingle}[~]~~,{}\<[E]%
\\
\>[B]{}\mathbf{type}\;\Conid{Dict0}\mathrel{=}\mbox{\textquotesingle}[~\mbox{\textquotesingle}(\text{\tt \char34 key\char34},\Conid{Bool})]~~,{}\<[E]%
\\
\>[B]{}\mathbf{type}\;\Conid{Dict1}\mathrel{=}\mbox{\textquotesingle}[~\mbox{\textquotesingle}(\text{\tt \char34 A\char34},\Conid{Int}),~\mbox{\textquotesingle}(\text{\tt \char34 B\char34},\text{\tt \char34 A\char34})]~~.{}\<[E]%
\ColumnHook
\end{hscode}\resethooks
All the entities defined above are types, where \ensuremath{\Conid{Dict0}} has kind \ensuremath{[\mskip1.5mu (\Conid{Symbol},\mathbin{*})\mskip1.5mu]}. In \ensuremath{\Conid{Dict1}}, while \ensuremath{\Conid{Int}} has kind \ensuremath{\mathbin{*}} and \ensuremath{\text{\tt \char34 A\char34}} has kind \ensuremath{\Conid{Symbol}}, the former kind subsumes the later. Thus \ensuremath{\Conid{Dict1}} also has kind \ensuremath{[\mskip1.5mu (\Conid{Symbol},\mathbin{*})\mskip1.5mu]}.

\subsection{Type-Level Functions}
\label{sec:type-fun}

Now that we can represent dictionaries as types, the next step is to define
operations on them. A function that inserts an entry to a dictionary, for
example, is a function from a type to a type. While it was shown that it is
possible to simulate type-level functions using Haskell type
classes~\cite{McBride:02:Faking}, in recent versions of GHC, {\em indexed type
families}, or \emph{type families} for short, are considered a cleaner solution.

For example, compare disjunction \ensuremath{(\mathrel{\vee})} and its type-level
counterpart \ensuremath{\Conid{Or}}:
\begin{hscode}\SaveRestoreHook
\column{B}{@{}>{\hspre}l<{\hspost}@{}}%
\column{7}{@{}>{\hspre}c<{\hspost}@{}}%
\column{7E}{@{}l@{}}%
\column{11}{@{}>{\hspre}l<{\hspost}@{}}%
\column{14}{@{}>{\hspre}l<{\hspost}@{}}%
\column{E}{@{}>{\hspre}l<{\hspost}@{}}%
\>[B]{}(\mathrel{\vee})\mathbin{::}\Conid{Bool}\to \Conid{Bool}\to \Conid{Bool}{}\<[E]%
\\
\>[B]{}\Conid{True}{}\<[7]%
\>[7]{}\mathrel{\vee}{}\<[7E]%
\>[11]{}\Varid{b}{}\<[14]%
\>[14]{}\mathrel{=}\Conid{True}{}\<[E]%
\\
\>[B]{}\Varid{a}{}\<[7]%
\>[7]{}\mathrel{\vee}{}\<[7E]%
\>[11]{}\Varid{b}{}\<[14]%
\>[14]{}\mathrel{=}\Varid{b}~~,{}\<[E]%
\ColumnHook
\end{hscode}\resethooks
\begin{hscode}\SaveRestoreHook
\column{B}{@{}>{\hspre}l<{\hspost}@{}}%
\column{3}{@{}>{\hspre}l<{\hspost}@{}}%
\column{10}{@{}>{\hspre}l<{\hspost}@{}}%
\column{16}{@{}>{\hspre}l<{\hspost}@{}}%
\column{24}{@{}>{\hspre}l<{\hspost}@{}}%
\column{E}{@{}>{\hspre}l<{\hspost}@{}}%
\>[B]{}\mathbf{type}\;\Varid{family}\;\Conid{Or}\;(\Varid{a}\mathbin{::}\Conid{Bool})\;(\Varid{b}\mathbin{::}\Conid{Bool})\mathbin{::}\Conid{Bool}{}\<[E]%
\\
\>[B]{}\hsindent{3}{}\<[3]%
\>[3]{}\mathbf{where}\;{}\<[10]%
\>[10]{}\mbox{\textquotesingle}\Conid{True}{}\<[16]%
\>[16]{}\mathrel{\Conid{`Or`}}\Varid{b}{}\<[24]%
\>[24]{}\mathrel{=}\mbox{\textquotesingle}\Conid{True}{}\<[E]%
\\
\>[10]{}\Varid{a}{}\<[16]%
\>[16]{}\mathrel{\Conid{`Or`}}\Varid{b}{}\<[24]%
\>[24]{}\mathrel{=}\Varid{b}~~.{}\<[E]%
\ColumnHook
\end{hscode}\resethooks
The first is a typical definition of \ensuremath{(\mathrel{\vee})} by pattern matching.
In the second definition, \ensuremath{\Conid{Bool}} is not a type, but a type lifted to a kind,
while \ensuremath{\Conid{True}} and \ensuremath{\Conid{False}} are types of kind \ensuremath{\Conid{Bool}}. The declaration says that
\ensuremath{\Conid{Or}} is a family of types, indexed by two parameters \ensuremath{\Varid{a}} and \ensuremath{\Varid{b}} of kind \ensuremath{\Conid{Bool}}.
The type with index \ensuremath{\mbox{\textquotesingle}\Conid{True}} and \ensuremath{\Varid{b}} is \ensuremath{\mbox{\textquotesingle}\Conid{True}}, and all other indices lead to \ensuremath{\Varid{b}}.
For our purpose, we can read \ensuremath{\Conid{Or}} as a function from types to types ---
observe how it resembles the term-level \ensuremath{(\mathrel{\vee})}. We present two more type-level
functions about \ensuremath{\Conid{Bool}} --- negation, and conditional, that we will use later:
\begin{hscode}\SaveRestoreHook
\column{B}{@{}>{\hspre}l<{\hspost}@{}}%
\column{3}{@{}>{\hspre}l<{\hspost}@{}}%
\column{14}{@{}>{\hspre}l<{\hspost}@{}}%
\column{E}{@{}>{\hspre}l<{\hspost}@{}}%
\>[B]{}\mathbf{type}\;\Varid{family}\;\Conid{Not}\;\Varid{a}\;\mathbf{where}{}\<[E]%
\\
\>[B]{}\hsindent{3}{}\<[3]%
\>[3]{}\Conid{Not}\;\mbox{\textquotesingle}\Conid{False}{}\<[14]%
\>[14]{}\mathrel{=}\mbox{\textquotesingle}\Conid{True}{}\<[E]%
\\
\>[B]{}\hsindent{3}{}\<[3]%
\>[3]{}\Conid{Not}\;\mbox{\textquotesingle}\Conid{True}{}\<[14]%
\>[14]{}\mathrel{=}\mbox{\textquotesingle}\Conid{False}~~,{}\<[E]%
\ColumnHook
\end{hscode}\resethooks
\begin{hscode}\SaveRestoreHook
\column{B}{@{}>{\hspre}l<{\hspost}@{}}%
\column{3}{@{}>{\hspre}l<{\hspost}@{}}%
\column{12}{@{}>{\hspre}l<{\hspost}@{}}%
\column{17}{@{}>{\hspre}l<{\hspost}@{}}%
\column{E}{@{}>{\hspre}l<{\hspost}@{}}%
\>[B]{}\mathbf{type}\;\Varid{family}\;\Conid{If}\;(\Varid{c}\mathbin{::}\Conid{Bool})\;(\Varid{t}\mathbin{::}\Varid{a})\;(\Varid{f}\mathbin{::}\Varid{a})\mathbin{::}\Varid{a}\;\mathbf{where}{}\<[E]%
\\
\>[B]{}\hsindent{3}{}\<[3]%
\>[3]{}\Conid{If}\;\mbox{\textquotesingle}\Conid{True}\;{}\<[12]%
\>[12]{}\Varid{tru}\;{}\<[17]%
\>[17]{}\Varid{fls}\mathrel{=}\Varid{tru}{}\<[E]%
\\
\>[B]{}\hsindent{3}{}\<[3]%
\>[3]{}\Conid{If}\;\mbox{\textquotesingle}\Conid{False}\;\Varid{tru}\;{}\<[17]%
\>[17]{}\Varid{fls}\mathrel{=}\Varid{fls}~~.{}\<[E]%
\ColumnHook
\end{hscode}\resethooks

As a remark, type families in GHC come in many flavors. One can define families of \ensuremath{\mathbf{data}} types, as well as families of \ensuremath{\mathbf{type}} synonyms. They can appear
inside type classes~\cite{tfclass,tfsynonym} or at toplevel. Top-level type families can be open~\cite{tfopen} or closed~\cite{tfclosed}. The flavor we
choose is top-level, closed type synonym families, since it allows overlapping
instances, and since we need none of the extensibility provided by open type
families. Notice that the instance \ensuremath{\mbox{\textquotesingle}\Conid{True}\mathrel{\Conid{`Or`}}\Varid{b}} could be subsumed under the
more general instance, \ensuremath{\Varid{a}\mathrel{\Conid{`Or`}}\Varid{b}}. In a closed type family we may resolve the
overlapping in order, just like how cases overlapping is resolved in term-level
functions.

Now we can define operations on type-level dictionaries. Let us begin
with:
\begin{hscode}\SaveRestoreHook
\column{B}{@{}>{\hspre}l<{\hspost}@{}}%
\column{5}{@{}>{\hspre}l<{\hspost}@{}}%
\column{20}{@{}>{\hspre}l<{\hspost}@{}}%
\column{33}{@{}>{\hspre}c<{\hspost}@{}}%
\column{33E}{@{}l@{}}%
\column{36}{@{}>{\hspre}l<{\hspost}@{}}%
\column{E}{@{}>{\hspre}l<{\hspost}@{}}%
\>[B]{}\mathbf{type}\;\Varid{family}\;\Conid{Get}\;(\Varid{xs}\mathbin{::}[\mskip1.5mu (\Conid{Symbol},\mathbin{*})\mskip1.5mu])\;(\Varid{k}\mathbin{::}\Conid{Symbol})\mathbin{::}\mathbin{*}\mathbf{where}{}\<[E]%
\\
\>[B]{}\hsindent{5}{}\<[5]%
\>[5]{}\Conid{Get}\;(\mbox{\textquotesingle}(\Varid{k},{}\<[20]%
\>[20]{}\Varid{x})\mathrel{\,\mbox{\textquotesingle}\!\!:}\Varid{xs})\;\Varid{k}{}\<[33]%
\>[33]{}\mathrel{=}{}\<[33E]%
\>[36]{}\Varid{x}{}\<[E]%
\\
\>[B]{}\hsindent{5}{}\<[5]%
\>[5]{}\Conid{Get}\;(\mbox{\textquotesingle}(\Varid{t},{}\<[20]%
\>[20]{}\Varid{x})\mathrel{\,\mbox{\textquotesingle}\!\!:}\Varid{xs})\;\Varid{k}{}\<[33]%
\>[33]{}\mathrel{=}{}\<[33E]%
\>[36]{}\Conid{Get}\;\Varid{xs}\;\Varid{k}~~.{}\<[E]%
\ColumnHook
\end{hscode}\resethooks
\ensuremath{\Conid{Get}\;\Varid{xs}\;\Varid{k}} returns the entry associated with key \ensuremath{\Varid{k}} in the
dictionary \ensuremath{\Varid{xs}}. Notice, in the first case, how type-level equality can be
expressed by unifying type variables with the same name. Note also that \ensuremath{\Conid{Get}} is
a partial function on types: while \ensuremath{\Conid{Get}\;\mbox{\textquotesingle}[\mbox{\textquotesingle}(\text{\tt \char34 A\char34},\Conid{Int})]\;\text{\tt \char34 A\char34}} evaluates
to \ensuremath{\Conid{Int}}, when \ensuremath{\Conid{Get}\;\mbox{\textquotesingle}[\mbox{\textquotesingle}(\text{\tt \char34 A\char34},\Conid{Int})]\;\text{\tt \char34 B\char34}} appears in a type expression,
there are no applicable rules to reduce it. The expression thus stays un-reduced.

%
Some other dictionary-related functions are defined
in Figure \ref{fig:dict-operations}. The function \ensuremath{\Conid{Set}} either updates an
existing entry or inserts a new entry, \ensuremath{\Conid{Del}} removes an entry matching
a given key, while \ensuremath{\Conid{Member}} checks whether a given key exists in the
dictionary.

\begin{figure}[t]
\begin{hscode}\SaveRestoreHook
\column{B}{@{}>{\hspre}l<{\hspost}@{}}%
\column{5}{@{}>{\hspre}l<{\hspost}@{}}%
\column{18}{@{}>{\hspre}l<{\hspost}@{}}%
\column{21}{@{}>{\hspre}l<{\hspost}@{}}%
\column{29}{@{}>{\hspre}l<{\hspost}@{}}%
\column{30}{@{}>{\hspre}l<{\hspost}@{}}%
\column{33}{@{}>{\hspre}l<{\hspost}@{}}%
\column{48}{@{}>{\hspre}l<{\hspost}@{}}%
\column{E}{@{}>{\hspre}l<{\hspost}@{}}%
\>[B]{}\mbox{\onelinecomment  inserts or updates an entry}{}\<[E]%
\\
\>[B]{}\mathbf{type}\;\Varid{family}\;\Conid{Set}\;{}\<[18]%
\>[18]{}(\Varid{xs}\mathbin{::}[\mskip1.5mu (\Conid{Symbol},\mathbin{*})\mskip1.5mu])\;(\Varid{k}\mathbin{::}\Conid{Symbol})\;(\Varid{x}\mathbin{::}\mathbin{*}){}\<[E]%
\\
\>[18]{}\mathbin{::}[\mskip1.5mu (\Conid{Symbol},\mathbin{*})\mskip1.5mu]\;\mathbf{where}{}\<[E]%
\\
\>[B]{}\hsindent{5}{}\<[5]%
\>[5]{}\Conid{Set}\;\mbox{\textquotesingle}[~]\;{}\<[29]%
\>[29]{}\Varid{k}\;\Varid{x}\mathrel{=}\mbox{\textquotesingle}[\mbox{\textquotesingle}(\Varid{k},\Varid{x})]{}\<[E]%
\\
\>[B]{}\hsindent{5}{}\<[5]%
\>[5]{}\Conid{Set}\;(\mbox{\textquotesingle}(\Varid{k},\Varid{y})\mathrel{\,\mbox{\textquotesingle}\!\!:}\Varid{xs})\;{}\<[29]%
\>[29]{}\Varid{k}\;\Varid{x}\mathrel{=}\mbox{\textquotesingle}(\Varid{k},\Varid{x}){}\<[48]%
\>[48]{}\mathrel{\,\mbox{\textquotesingle}\!\!:}\Varid{xs}{}\<[E]%
\\
\>[B]{}\hsindent{5}{}\<[5]%
\>[5]{}\Conid{Set}\;(\mbox{\textquotesingle}(\Varid{t},\Varid{y})\mathrel{\,\mbox{\textquotesingle}\!\!:}\Varid{xs})\;{}\<[29]%
\>[29]{}\Varid{k}\;\Varid{x}\mathrel{=}\mbox{\textquotesingle}(\Varid{t},\Varid{y}){}\<[48]%
\>[48]{}\mathrel{\,\mbox{\textquotesingle}\!\!:}\Conid{Set}\;\Varid{xs}\;\Varid{k}\;\Varid{x}{}\<[E]%
\\[\blanklineskip]%
\>[B]{}\mbox{\onelinecomment  removes an entry}{}\<[E]%
\\
\>[B]{}\mathbf{type}\;\Varid{family}\;\Conid{Del}\;{}\<[18]%
\>[18]{}(\Varid{xs}\mathbin{::}[\mskip1.5mu (\Conid{Symbol},\mathbin{*})\mskip1.5mu])\;(\Varid{k}\mathbin{::}\Conid{Symbol}){}\<[E]%
\\
\>[18]{}\mathbin{::}[\mskip1.5mu (\Conid{Symbol},\mathbin{*})\mskip1.5mu]\;\mathbf{where}{}\<[E]%
\\
\>[B]{}\hsindent{5}{}\<[5]%
\>[5]{}\Conid{Del}\;\mbox{\textquotesingle}[~]\;{}\<[30]%
\>[30]{}\Varid{k}\mathrel{=}\mbox{\textquotesingle}[~]{}\<[E]%
\\
\>[B]{}\hsindent{5}{}\<[5]%
\>[5]{}\Conid{Del}\;(\mbox{\textquotesingle}(\Varid{k},\Varid{y})\mathrel{\,\mbox{\textquotesingle}\!\!:}\Varid{xs})\;{}\<[30]%
\>[30]{}\Varid{k}\mathrel{=}\Varid{xs}{}\<[E]%
\\
\>[B]{}\hsindent{5}{}\<[5]%
\>[5]{}\Conid{Del}\;(\mbox{\textquotesingle}(\Varid{t},\Varid{y})\mathrel{\,\mbox{\textquotesingle}\!\!:}\Varid{xs})\;{}\<[30]%
\>[30]{}\Varid{k}\mathrel{=}\mbox{\textquotesingle}(\Varid{t},\Varid{y})\mathrel{\,\mbox{\textquotesingle}\!\!:}\Conid{Del}\;\Varid{xs}\;\Varid{k}{}\<[E]%
\\[\blanklineskip]%
\>[B]{}\mbox{\onelinecomment  membership}{}\<[E]%
\\
\>[B]{}\mathbf{type}\;\Varid{family}\;\Conid{Member}\;{}\<[21]%
\>[21]{}(\Varid{xs}\mathbin{::}[\mskip1.5mu (\Conid{Symbol},\mathbin{*})\mskip1.5mu])\;(\Varid{k}\mathbin{::}\Conid{Symbol}){}\<[E]%
\\
\>[21]{}\mathbin{::}\Conid{Bool}\;\mathbf{where}{}\<[E]%
\\
\>[B]{}\hsindent{5}{}\<[5]%
\>[5]{}\Conid{Member}\;\mbox{\textquotesingle}[~]\;{}\<[33]%
\>[33]{}\Varid{k}\mathrel{=}\mbox{\textquotesingle}\Conid{False}{}\<[E]%
\\
\>[B]{}\hsindent{5}{}\<[5]%
\>[5]{}\Conid{Member}\;(\mbox{\textquotesingle}(\Varid{k},\Varid{x})\mathrel{\,\mbox{\textquotesingle}\!\!:}\Varid{xs})\;{}\<[33]%
\>[33]{}\Varid{k}\mathrel{=}\mbox{\textquotesingle}\Conid{True}{}\<[E]%
\\
\>[B]{}\hsindent{5}{}\<[5]%
\>[5]{}\Conid{Member}\;(\mbox{\textquotesingle}(\Varid{t},\Varid{x})\mathrel{\,\mbox{\textquotesingle}\!\!:}\Varid{xs})\;{}\<[33]%
\>[33]{}\Varid{k}\mathrel{=}\Conid{Member}\;\Varid{xs}\;\Varid{k}{}\<[E]%
\\
\>[B]{}\hsindent{5}{}\<[5]%
\>[5]{}~~{}\<[E]%
\ColumnHook
\end{hscode}\resethooks
\vspace{-1cm}
\caption{Some operations on type-level dictionaries.}
\label{fig:dict-operations}
\end{figure}

%% file: sections/Embedding.tex
%
%
\makeatletter
\@ifundefined{lhs2tex.lhs2tex.sty.read}%
  {\@namedef{lhs2tex.lhs2tex.sty.read}{}%
   \newcommand\SkipToFmtEnd{}%
   \newcommand\EndFmtInput{}%
   \long\def\SkipToFmtEnd#1\EndFmtInput{}%
  }\SkipToFmtEnd

\newcommand\ReadOnlyOnce[1]{\@ifundefined{#1}{\@namedef{#1}{}}\SkipToFmtEnd}
\usepackage{amstext}
\usepackage{amssymb}
\usepackage{stmaryrd}
\DeclareFontFamily{OT1}{cmtex}{}
\DeclareFontShape{OT1}{cmtex}{m}{n}
  {<5><6><7><8>cmtex8
   <9>cmtex9
   <10><10.95><12><14.4><17.28><20.74><24.88>cmtex10}{}
\DeclareFontShape{OT1}{cmtex}{m}{it}
  {<-> ssub * cmtt/m/it}{}
\newcommand{\texfamily}{\fontfamily{cmtex}\selectfont}
\DeclareFontShape{OT1}{cmtt}{bx}{n}
  {<5><6><7><8>cmtt8
   <9>cmbtt9
   <10><10.95><12><14.4><17.28><20.74><24.88>cmbtt10}{}
\DeclareFontShape{OT1}{cmtex}{bx}{n}
  {<-> ssub * cmtt/bx/n}{}
\newcommand{\tex}[1]{\text{\texfamily#1}}	

\newcommand{\Sp}{\hskip.33334em\relax}

\newcommand{\Conid}[1]{\mathit{#1}}
\newcommand{\Varid}[1]{\mathit{#1}}
\newcommand{\anonymous}{\kern0.06em \vbox{\hrule\@width.5em}}
\newcommand{\plus}{\mathbin{+\!\!\!+}}
\newcommand{\bind}{\mathbin{>\!\!\!>\mkern-6.7mu=}}
\newcommand{\rbind}{\mathbin{=\mkern-6.7mu<\!\!\!<}}
\newcommand{\sequ}{\mathbin{>\!\!\!>}}
\renewcommand{\leq}{\leqslant}
\renewcommand{\geq}{\geqslant}
\usepackage{polytable}

\@ifundefined{mathindent}%
  {\newdimen\mathindent\mathindent\leftmargini}%
  {}%

\def\resethooks{%
  \global\let\SaveRestoreHook\empty
  \global\let\ColumnHook\empty}
\newcommand*{\savecolumns}[1][default]%
  {\g@addto@macro\SaveRestoreHook{\savecolumns[#1]}}
\newcommand*{\restorecolumns}[1][default]%
  {\g@addto@macro\SaveRestoreHook{\restorecolumns[#1]}}
\newcommand*{\aligncolumn}[2]%
  {\g@addto@macro\ColumnHook{\column{#1}{#2}}}

\resethooks

\newcommand{\onelinecommentchars}{\quad-{}- }
\newcommand{\commentbeginchars}{\enskip\{-}
\newcommand{\commentendchars}{-\}\enskip}

\newcommand{\visiblecomments}{%
  \let\onelinecomment=\onelinecommentchars
  \let\commentbegin=\commentbeginchars
  \let\commentend=\commentendchars}

\newcommand{\invisiblecomments}{%
  \let\onelinecomment=\empty
  \let\commentbegin=\empty
  \let\commentend=\empty}

\visiblecomments

\newlength{\blanklineskip}
\setlength{\blanklineskip}{0.66084ex}

\newcommand{\hsindent}[1]{\quad}
\let\hspre\empty
\let\hspost\empty
\newcommand{\NB}{\textbf{NB}}
\newcommand{\Todo}[1]{$\langle$\textbf{To do:}~#1$\rangle$}

\EndFmtInput
\makeatother
%
%
%
%
%
%
%
%
%
\ReadOnlyOnce{polycode.fmt}%
\makeatletter

\newcommand{\hsnewpar}[1]%
  {{\parskip=0pt\parindent=0pt\par\vskip #1\noindent}}

\newcommand{\hscodestyle}{}


\newcommand{\sethscode}[1]%
  {\expandafter\let\expandafter\hscode\csname #1\endcsname
   \expandafter\let\expandafter\endhscode\csname end#1\endcsname}


\newenvironment{compathscode}%
  {\par\noindent
   \advance\leftskip\mathindent
   \hscodestyle
   \let\\=\@normalcr
   \let\hspre\(\let\hspost\)%
   \pboxed}%
  {\endpboxed\)%
   \par\noindent
   \ignorespacesafterend}

\newcommand{\compaths}{\sethscode{compathscode}}


\newenvironment{plainhscode}%
  {\hsnewpar\abovedisplayskip
   \advance\leftskip\mathindent
   \hscodestyle
   \let\hspre\(\let\hspost\)%
   \pboxed}%
  {\endpboxed%
   \hsnewpar\belowdisplayskip
   \ignorespacesafterend}

\newenvironment{oldplainhscode}%
  {\hsnewpar\abovedisplayskip
   \advance\leftskip\mathindent
   \hscodestyle
   \let\\=\@normalcr
   \(\pboxed}%
  {\endpboxed\)%
   \hsnewpar\belowdisplayskip
   \ignorespacesafterend}


\newcommand{\plainhs}{\sethscode{plainhscode}}
\newcommand{\oldplainhs}{\sethscode{oldplainhscode}}
\plainhs


\newenvironment{arrayhscode}%
  {\hsnewpar\abovedisplayskip
   \advance\leftskip\mathindent
   \hscodestyle
   \let\\=\@normalcr
   \(\parray}%
  {\endparray\)%
   \hsnewpar\belowdisplayskip
   \ignorespacesafterend}

\newcommand{\arrayhs}{\sethscode{arrayhscode}}


\newenvironment{mathhscode}%
  {\parray}{\endparray}

\newcommand{\mathhs}{\sethscode{mathhscode}}


\newenvironment{texthscode}%
  {\(\parray}{\endparray\)}

\newcommand{\texths}{\sethscode{texthscode}}


\def\codeframewidth{\arrayrulewidth}
\RequirePackage{calc}

\newenvironment{framedhscode}%
  {\parskip=\abovedisplayskip\par\noindent
   \hscodestyle
   \arrayrulewidth=\codeframewidth
   \tabular{@{}|p{\linewidth-2\arraycolsep-2\arrayrulewidth-2pt}|@{}}%
   \hline\framedhslinecorrect\\{-1.5ex}%
   \let\endoflinesave=\\
   \let\\=\@normalcr
   \(\pboxed}%
  {\endpboxed\)%
   \framedhslinecorrect\endoflinesave{.5ex}\hline
   \endtabular
   \parskip=\belowdisplayskip\par\noindent
   \ignorespacesafterend}

\newcommand{\framedhslinecorrect}[2]%
  {#1[#2]}

\newcommand{\framedhs}{\sethscode{framedhscode}}


\newenvironment{inlinehscode}%
  {\(\def\column##1##2{}%
   \let\>\undefined\let\<\undefined\let\\\undefined
   \newcommand\>[1][]{}\newcommand\<[1][]{}\newcommand\\[1][]{}%
   \def\fromto##1##2##3{##3}%
   \def\nextline{}}{\) }%

\newcommand{\inlinehs}{\sethscode{inlinehscode}}


\newenvironment{joincode}%
  {\let\orighscode=\hscode
   \let\origendhscode=\endhscode
   \def\endhscode{\def\hscode{\endgroup\def\@currenvir{hscode}\\}\begingroup}
   \orighscode\def\hscode{\endgroup\def\@currenvir{hscode}}}%
  {\origendhscode
   \global\let\hscode=\orighscode
   \global\let\endhscode=\origendhscode}%

\makeatother
\EndFmtInput

\ReadOnlyOnce{Formatting.fmt}%
\makeatletter

\let\Varid\mathit
\let\Conid\mathsf

\def\commentbegin{\quad\{\ }
\def\commentend{\}}

\newcommand{\ty}[1]{\Conid{#1}}
\newcommand{\con}[1]{\Conid{#1}}
\newcommand{\id}[1]{\Varid{#1}}
\newcommand{\cl}[1]{\Varid{#1}}
\newcommand{\opsym}[1]{\mathrel{#1}}

\newcommand\Keyword[1]{\textbf{\textsf{#1}}}
\newcommand\Hide{\mathbin{\downarrow}}
\newcommand\Reveal{\mathbin{\uparrow}}


\makeatother
\EndFmtInput

\section{Embedding \Hedis{} Commands}
\label{sec:embedding-commands}

Having the indexed monads and type-level dictionaries, in this section we
present our embedding of \Hedis{} commands into \Edis{}, while introducing
necessary concepts when they are used.

\subsection{Proxies and Singleton Types}
\label{sec:proxy-key}

The \Hedis{} function \ensuremath{\Varid{del}\mathbin{::}[\mskip1.5mu \Conid{ByteString}\mskip1.5mu]\to \Conid{Redis}\;(\Conid{R} \uplus \Conid{Integer})} takes a list of keys (encoded to \ensuremath{\Conid{ByteString}}) and removes the entries having those
keys in the database. For some reason to be explained later, we consider an \Edis{}
counterpart that takes only one key. A first attempt may lead to something like
the following:
\begin{hscode}\SaveRestoreHook
\column{B}{@{}>{\hspre}l<{\hspost}@{}}%
\column{E}{@{}>{\hspre}l<{\hspost}@{}}%
\>[B]{}\Varid{del}\mathbin{::}\Conid{String}\to \Conid{Edis}\;\Varid{xs}\;(\Conid{Del}\;\Varid{xs}~\mathbin{?})\;(\Conid{R} \uplus \Conid{Integer}){}\<[E]%
\\
\>[B]{}\Varid{del}\;\Varid{key}\mathrel{=}\Conid{Edis}\;(\Varid{\Conid{Hedis}.del}\;[\mskip1.5mu \Varid{encode}\;\Varid{key}\mskip1.5mu])~~,{}\<[E]%
\ColumnHook
\end{hscode}\resethooks
where the function \ensuremath{\Varid{encode}} converts \ensuremath{\Conid{String}} to \ensuremath{\Conid{ByteString}}. At term-level,
our \ensuremath{\Varid{del}} merely calls \ensuremath{\Varid{\Conid{Hedis}.del}}. At type-level, if the status of the database
before \ensuremath{\Varid{del}} is called meets the constraint represented by the dictionary
\ensuremath{\Varid{xs}}, the status afterwards should meet the constraint \ensuremath{\Conid{Del}\;\Varid{xs}~\mathbin{?}}. The question, however, is what to fill in place of the question mark. It cannot be
\ensuremath{\Conid{Del}\;\Varid{xs}\;\Varid{key}}, since \ensuremath{\Varid{key}} is a runtime value and not a type. How do we smuggle
a runtime value to type-level?

In a language with phase distinction like Haskell, it is certainly impossible
to pass the value of \ensuremath{\Varid{key}} to the type checker if it truly is a runtime value,
for example, a string read from the user. If the value of \ensuremath{\Varid{key}} can be
determined statically, however, {\em singleton types}~\cite{singletons} can be used to represent a
type as a value, thus build a connection between the two realms.

A singleton type is a type that has only one term. When the term is built, it
carries a type that can be inspected by the type checker. The term can be
thought of as a representative of its type at the realm of runtime values. For
our purpose, we will use the following type \ensuremath{\Conid{Proxy}}:
\begin{hscode}\SaveRestoreHook
\column{B}{@{}>{\hspre}l<{\hspost}@{}}%
\column{E}{@{}>{\hspre}l<{\hspost}@{}}%
\>[B]{}\mathbf{data}\;\Conid{Proxy}\;\Varid{t}\mathrel{=}\Conid{Proxy}~~.{}\<[E]%
\ColumnHook
\end{hscode}\resethooks
For every type \ensuremath{\Varid{t}}, \ensuremath{\Conid{Proxy}\;\Varid{t}} is a type that has only one term: \ensuremath{\Conid{Proxy}}.%
\footnote{While giving the same name to both the type and the term can be very
confusing, it is unfortunately a common practice in the Haskell community.}
To call \ensuremath{\Varid{del}}, instead of passing a key as a \ensuremath{\Conid{String}} value, we give it a proxy
with a specified type:
\begin{hscode}\SaveRestoreHook
\column{B}{@{}>{\hspre}l<{\hspost}@{}}%
\column{E}{@{}>{\hspre}l<{\hspost}@{}}%
\>[B]{}\Varid{del}\;(\Conid{Proxy}\mathbin{::}\Conid{Proxy}\;\text{\tt \char34 A\char34})~~,{}\<[E]%
\ColumnHook
\end{hscode}\resethooks
where \ensuremath{\text{\tt \char34 A\char34}} is not a value, but a string lifted to a type (of kind \ensuremath{\Conid{Symbol}}).
Now that the type checker has access to the key, the type of \ensuremath{\Varid{del}} can be
\ensuremath{\Conid{Proxy}\;\Varid{k}\to \Conid{Edis}\;\Varid{xs}\;(\Conid{Del}\;\Varid{xs}\;\Varid{k})\;(\Conid{R} \uplus \Conid{Integer})}.

The next problem is that, \ensuremath{\Varid{del}}, at term level, gets only a value constructor
\ensuremath{\Conid{Proxy}} without further information, while it needs to pass a \ensuremath{\Conid{ByteString}} key
to \ensuremath{\Varid{\Conid{Hedis}.del}}. Every concrete string literal lifted to a type, for example,
\ensuremath{\text{\tt \char34 A\char34}}, belongs to a type class \ensuremath{\Conid{KnownSymbol}}. For all types \ensuremath{\Varid{k}} in \ensuremath{\Conid{KnownSymbol}},
the function \ensuremath{\Varid{symbolVal}}:\begin{hscode}\SaveRestoreHook
\column{B}{@{}>{\hspre}l<{\hspost}@{}}%
\column{3}{@{}>{\hspre}l<{\hspost}@{}}%
\column{E}{@{}>{\hspre}l<{\hspost}@{}}%
\>[3]{}\Varid{symbolVal}\mathbin{::}\Conid{KnownSymbol}\;\Varid{k}\Rightarrow \Varid{proxy}\;\Varid{k}\to \Conid{String}~~,{}\<[E]%
\ColumnHook
\end{hscode}\resethooks
retrieves the string associated with a type-level literal that is known at
compile time. In summary, \ensuremath{\Varid{del}} can be implemented as:
\begin{hscode}\SaveRestoreHook
\column{B}{@{}>{\hspre}l<{\hspost}@{}}%
\column{9}{@{}>{\hspre}l<{\hspost}@{}}%
\column{45}{@{}>{\hspre}l<{\hspost}@{}}%
\column{E}{@{}>{\hspre}l<{\hspost}@{}}%
\>[B]{}\Varid{del}\mathbin{::}{}\<[9]%
\>[9]{}\Conid{KnownSymbol}\;\Varid{k}\Rightarrow {}\<[E]%
\\
\>[9]{}\Conid{Proxy}\;\Varid{k}\to \Conid{Edis}\;\Varid{xs}\;(\Conid{Del}\;\Varid{xs}\;\Varid{k})\;(\Conid{R} \uplus \Conid{Integer}){}\<[E]%
\\
\>[B]{}\Varid{del}\;\Varid{key}\mathrel{=}\Conid{Edis}\;(\Varid{\Conid{Hedis}.del}\;[\mskip1.5mu \Varid{encodeKey}\;\Varid{key}\mskip1.5mu]){}\<[45]%
\>[45]{}~~,{}\<[E]%
\ColumnHook
\end{hscode}\resethooks
where \ensuremath{\Varid{encodeKey}\mathrel{=}\Varid{encode}\mathbin{\cdot}\Varid{symbolVal}}.

A final note: the function \ensuremath{\Varid{encode}}, from the Haskell library {\sc cereal},
helps to convert certain datatypes that are {\em serializable} into \ensuremath{\Conid{ByteString}}.
The function and its dual \ensuremath{\Varid{decode}} will be used more later.
\begin{hscode}\SaveRestoreHook
\column{B}{@{}>{\hspre}l<{\hspost}@{}}%
\column{9}{@{}>{\hspre}l<{\hspost}@{}}%
\column{E}{@{}>{\hspre}l<{\hspost}@{}}%
\>[B]{}\Varid{encode}{}\<[9]%
\>[9]{}\mathbin{::}\Conid{Serialize}\;\Varid{a}\Rightarrow \Varid{a}\to \Conid{ByteString}~~,{}\<[E]%
\\
\>[B]{}\Varid{decode}{}\<[9]%
\>[9]{}\mathbin{::}\Conid{Serialize}\;\Varid{a}\Rightarrow \Conid{ByteString}\to \Conid{Either}\;\Conid{String}\;\Varid{a}~~.{}\<[E]%
\ColumnHook
\end{hscode}\resethooks

\subsection{Automatic Serialization}
\label{sec:polymorphic-redis}

As mentioned before, while \Redis{} provide a number of container types
including lists, sets, and hash, etc., the primitive type is string. \Hedis{}
programmers manually convert data of other types to strings before saving them
into the data store. In \Edis{}, we wish to save some of the effort for the
programmers, as well as keeping a careful record of the intended types of the
strings in the data store.

To keep track of intended types of strings in the data store, we define the
following types (that have no terms):
\begin{hscode}\SaveRestoreHook
\column{B}{@{}>{\hspre}l<{\hspost}@{}}%
\column{16}{@{}>{\hspre}l<{\hspost}@{}}%
\column{E}{@{}>{\hspre}l<{\hspost}@{}}%
\>[B]{}\mathbf{data}\;\Conid{StringOf}{}\<[16]%
\>[16]{}\mathbin{::}\mathbin{*}\to \mathbin{*}~~,{}\<[E]%
\\
\>[B]{}\mathbf{data}\;\Conid{ListOf}{}\<[16]%
\>[16]{}\mathbin{::}\mathbin{*}\to \mathbin{*}~~,{}\<[E]%
\\
\>[B]{}\mathbf{data}\;\Conid{SetOf}{}\<[16]%
\>[16]{}\mathbin{::}\mathbin{*}\to \mathbin{*}~~...{}\<[E]%
\ColumnHook
\end{hscode}\resethooks
If a key is associated with, for example, \ensuremath{\Conid{StringOf}\;\Conid{Int}} in our dictionary, we
mean that its value in the data store was serialized from an \ensuremath{\Conid{Int}} and should be
used as an \ensuremath{\Conid{Int}}. Types \ensuremath{\Conid{ListOf}\;\Varid{a}} and \ensuremath{\Conid{SetOf}\;\Varid{a}}, respectively, denotes that the
value is a list or a set of type \ensuremath{\Varid{a}}.

While the \ensuremath{\Varid{set}} command in \Hedis{} always writes a string to the data store,
the corresponding \ensuremath{\Varid{set}} in \Redis{} applies to any serializable type (those
in the class \ensuremath{\Conid{Serialize}}), and performs the encoding for the user:
\begin{hscode}\SaveRestoreHook
\column{B}{@{}>{\hspre}l<{\hspost}@{}}%
\column{9}{@{}>{\hspre}l<{\hspost}@{}}%
\column{11}{@{}>{\hspre}l<{\hspost}@{}}%
\column{E}{@{}>{\hspre}l<{\hspost}@{}}%
\>[B]{}\Varid{set}\mathbin{::}{}\<[9]%
\>[9]{}(\Conid{KnownSymbol}\;\Varid{k},\Conid{Serialize}\;\Varid{a})\Rightarrow \Conid{Proxy}\;\Varid{k}\to \Varid{a}\to {}\<[E]%
\\
\>[9]{}\hsindent{2}{}\<[11]%
\>[11]{}\Conid{Edis}\;\Varid{xs}\;(\Conid{Set}\;\Varid{xs}\;\Varid{k}\;(\Conid{StringOf}\;\Varid{a}))\;(\Conid{Either}\;\Conid{Reply}\;\Conid{Status}){}\<[E]%
\\
\>[B]{}\Varid{set}\;\Varid{key}\;\Varid{v}\mathrel{=}\Conid{Edis}\;(\Varid{\Conid{Hedis}.set}\;(\Varid{encodeKey}\;\Varid{key})\;(\Varid{encode}\;\Varid{v}))~~,{}\<[E]%
\ColumnHook
\end{hscode}\resethooks
For example, executing \ensuremath{\Varid{set}\;(\Conid{Proxy}\mathbin{::}\Conid{Proxy}\;\text{\tt \char34 A\char34})\;\Conid{True}} updates the dictionary
with an entry \ensuremath{\mbox{\textquotesingle}(\text{\tt \char34 A\char34},\Conid{StringOf}\;\Conid{Bool})}. If \ensuremath{\text{\tt \char34 A\char34}} is not in the dictionary,
this entry is added; otherwise the old type of \ensuremath{\text{\tt \char34 A\char34}} is updated to
\ensuremath{\Conid{StringOf}\;\Conid{Bool}}.

\Redis{} command \texttt{INCR} reads the (string) value of the given key, parses
it as an integer, and increments it by one, before storing it back. The command
\texttt{INCRBYFLOAT} increments the floating point value of a key by a given
amount. They are defined in \Edis{} below:
\begin{hscode}\SaveRestoreHook
\column{B}{@{}>{\hspre}l<{\hspost}@{}}%
\column{3}{@{}>{\hspre}l<{\hspost}@{}}%
\column{10}{@{}>{\hspre}l<{\hspost}@{}}%
\column{11}{@{}>{\hspre}l<{\hspost}@{}}%
\column{E}{@{}>{\hspre}l<{\hspost}@{}}%
\>[B]{}\Varid{incr}\mathbin{::}{}\<[10]%
\>[10]{}(\Conid{KnownSymbol}\;\Varid{k},\Conid{Get}\;\Varid{xs}\;\Varid{k}\mathrel{\sim}\Conid{StringOf}\;\Conid{Integer})\Rightarrow {}\<[E]%
\\
\>[10]{}\Conid{Proxy}\;\Varid{k}\to \Conid{Edis}\;\Varid{xs}\;\Varid{xs}\;(\Conid{R} \uplus \Conid{Integer}){}\<[E]%
\\
\>[B]{}\Varid{incr}\;\Varid{key}\mathrel{=}\Conid{Edis}\;(\Varid{\Conid{Hedis}.incr}\;(\Varid{encodeKey}\;\Varid{key}))~~,{}\<[E]%
\\[\blanklineskip]%
\>[B]{}\Varid{incrbyfloat}\mathbin{::}(\Conid{KnownSymbol}\;\Varid{k},\Conid{Get}\;\Varid{xs}\;\Varid{k}\mathrel{\sim}\Conid{StringOf}\;\Conid{Double}){}\<[E]%
\\
\>[B]{}\hsindent{11}{}\<[11]%
\>[11]{}\Rightarrow \Conid{Proxy}\;\Varid{k}\to \Conid{Double}\to \Conid{Edis}\;\Varid{xs}\;\Varid{xs}\;(\Conid{R} \uplus \Conid{Double}){}\<[E]%
\\
\>[B]{}\Varid{incrbyfloat}\;\Varid{key}\;\Varid{eps}\mathrel{=}{}\<[E]%
\\
\>[B]{}\hsindent{3}{}\<[3]%
\>[3]{}\Conid{Edis}\;(\Varid{\Conid{Hedis}.incrbyfloat}\;(\Varid{encodeKey}\;\Varid{key})\;\Varid{eps})~~.{}\<[E]%
\ColumnHook
\end{hscode}\resethooks
Notice the use of (\ensuremath{\mathrel{\sim}}), \emph{equality constraints}~\cite{typeeq}, to enforce
that the intended type of the value of \ensuremath{\Varid{k}} must respectively be \ensuremath{\Conid{Integer}} and
\ensuremath{\Conid{Double}}. The function \ensuremath{\Varid{incr}} is only allowed to be called in a context where
the type checker is able to reduce \ensuremath{\Conid{Get}\;\Varid{xs}\;\Varid{k}} to \ensuremath{\Conid{StringOf}\;\Conid{Integer}} ---
recall that when \ensuremath{\Varid{k}} is not in \ensuremath{\Varid{xs}}, \ensuremath{\Conid{Get}\;\Varid{xs}\;\Varid{k}} cannot be fully reduced. The
type of \ensuremath{\Varid{incrbyfloat}} works in a similar way.

\subsection{Disjunctive Constraints}
\label{sec:disjunctive-constraints}

Recall, from Section \ref{sec:introduction}, that commands \texttt{LPUSH key
val} and \texttt{LLEN key} succeed either when \ensuremath{\Varid{key}} appears in the
data store and is assigned a list, or when \ensuremath{\Varid{key}} does not appear at all.
What we wish to have in their constraint is thus a predicate equivalent to \ensuremath{\Conid{Get}\;\Varid{xs}\;\Varid{k}\doubleequals\Conid{ListOf}\;\Varid{a}\mathrel{\vee}\neg \;(\Conid{Member}\;\Varid{xs}\;\Varid{k})}. \hide{In fact, many \Redis{} commands
are invokable under such ``well-typed, or non-existent'' precondition.}

To impose a conjunctive constraint \ensuremath{\Conid{P}\mathrel{\wedge}\Conid{Q}}, one may simply put them both in the
type: \ensuremath{(\Conid{P},\Conid{Q})\Rightarrow \mathbin{...}}. Expressing disjunctive constraints is only slightly
harder, thanks to our type-level functions. We may thus write the predicate as:
\begin{hscode}\SaveRestoreHook
\column{B}{@{}>{\hspre}l<{\hspost}@{}}%
\column{E}{@{}>{\hspre}l<{\hspost}@{}}%
\>[B]{}\Conid{Get}\;\Varid{xs}\;\Varid{k}\mathrel{\sim}\Conid{ListOf}\;\Varid{a}\mathrel{\Conid{`Or`}}\Conid{Not}\;(\Conid{Member}\;\Varid{xs}\;\Varid{k})~~.{}\<[E]%
\ColumnHook
\end{hscode}\resethooks
To avoid referring to \ensuremath{\Varid{a}}, which might not exist, we define an auxiliary predicate \ensuremath{\Conid{IsList}\mathbin{::}\mathbin{*}\to \Conid{Bool}} such that \ensuremath{\Conid{IsList}\;\Varid{t}} reduces to \ensuremath{\mbox{\textquotesingle}\Conid{True}}
only if \ensuremath{\Varid{t}\mathrel{=}\Conid{ListOf}\;\Varid{a}}. As many \Redis{} commands are invokable only under such
``well-typed, or non-existent'' precondition, we give names to such constraints,
as seen in Figure~\ref{fig:xxxOrNX}.\\

\begin{figure}[t]
\begin{hscode}\SaveRestoreHook
\column{B}{@{}>{\hspre}l<{\hspost}@{}}%
\column{3}{@{}>{\hspre}l<{\hspost}@{}}%
\column{5}{@{}>{\hspre}l<{\hspost}@{}}%
\column{14}{@{}>{\hspre}l<{\hspost}@{}}%
\column{18}{@{}>{\hspre}l<{\hspost}@{}}%
\column{22}{@{}>{\hspre}l<{\hspost}@{}}%
\column{24}{@{}>{\hspre}l<{\hspost}@{}}%
\column{28}{@{}>{\hspre}l<{\hspost}@{}}%
\column{E}{@{}>{\hspre}l<{\hspost}@{}}%
\>[B]{}\mathbf{type}\;\Varid{family}\;\Conid{IsList}\;(\Varid{t}\mathbin{::}\mathbin{*})\mathbin{::}\Conid{Bool}\;\mathbf{where}{}\<[E]%
\\
\>[B]{}\hsindent{5}{}\<[5]%
\>[5]{}\Conid{IsList}\;(\Conid{ListOf}\;\Varid{a}){}\<[24]%
\>[24]{}\mathrel{=}\mbox{\textquotesingle}\Conid{True}{}\<[E]%
\\
\>[B]{}\hsindent{5}{}\<[5]%
\>[5]{}\Conid{IsList}\;\Varid{t}{}\<[24]%
\>[24]{}\mathrel{=}\mbox{\textquotesingle}\Conid{False}{}\<[E]%
\\
\>[B]{}\mathbf{type}\;\Varid{family}\;\Conid{IsSet}\;(\Varid{t}\mathbin{::}\mathbin{*})\mathbin{::}\Conid{Bool}\;\mathbf{where}{}\<[E]%
\\
\>[B]{}\hsindent{5}{}\<[5]%
\>[5]{}\Conid{IsSet}\;(\Conid{SetOf}\;\Varid{a}){}\<[22]%
\>[22]{}\mathrel{=}\mbox{\textquotesingle}\Conid{True}{}\<[E]%
\\
\>[B]{}\hsindent{5}{}\<[5]%
\>[5]{}\Conid{IsSet}\;\Varid{t}{}\<[22]%
\>[22]{}\mathrel{=}\mbox{\textquotesingle}\Conid{False}{}\<[E]%
\\
\>[B]{}\mathbf{type}\;\Varid{family}\;\Conid{IsString}\;(\Varid{t}\mathbin{::}\mathbin{*})\mathbin{::}\Conid{Bool}\;\mathbf{where}{}\<[E]%
\\
\>[B]{}\hsindent{5}{}\<[5]%
\>[5]{}\Conid{IsString}\;(\Conid{StringOf}\;\Varid{a}){}\<[28]%
\>[28]{}\mathrel{=}\mbox{\textquotesingle}\Conid{True}{}\<[E]%
\\
\>[B]{}\hsindent{5}{}\<[5]%
\>[5]{}\Conid{IsString}\;\Varid{t}{}\<[28]%
\>[28]{}\mathrel{=}\mbox{\textquotesingle}\Conid{False}{}\<[E]%
\\[\blanklineskip]%
\>[B]{}\mathbf{type}\;\Conid{ListOrNX}\;{}\<[18]%
\>[18]{}\Varid{xs}\;\Varid{k}\mathrel{=}{}\<[E]%
\\
\>[B]{}\hsindent{3}{}\<[3]%
\>[3]{}(\Conid{IsList}\;{}\<[14]%
\>[14]{}(\Conid{Get}\;\Varid{xs}\;\Varid{k})\mathrel{\Conid{`Or`}}\Conid{Not}\;(\Conid{Member}\;\Varid{xs}\;\Varid{k}))\mathrel{\sim}\mbox{\textquotesingle}\Conid{True}{}\<[E]%
\\
\>[B]{}\mathbf{type}\;\Conid{SetOrNX}\;{}\<[18]%
\>[18]{}\Varid{xs}\;\Varid{k}\mathrel{=}{}\<[E]%
\\
\>[B]{}\hsindent{3}{}\<[3]%
\>[3]{}(\Conid{IsSet}\;{}\<[14]%
\>[14]{}(\Conid{Get}\;\Varid{xs}\;\Varid{k})\mathrel{\Conid{`Or`}}\Conid{Not}\;(\Conid{Member}\;\Varid{xs}\;\Varid{k}))\mathrel{\sim}\mbox{\textquotesingle}\Conid{True}{}\<[E]%
\\
\>[B]{}\mathbf{type}\;\Conid{StringOrNX}\;{}\<[18]%
\>[18]{}\Varid{xs}\;\Varid{k}\mathrel{=}{}\<[E]%
\\
\>[B]{}\hsindent{3}{}\<[3]%
\>[3]{}(\Conid{IsString}\;{}\<[14]%
\>[14]{}(\Conid{Get}\;\Varid{xs}\;\Varid{k})\mathrel{\Conid{`Or`}}\Conid{Not}\;(\Conid{Member}\;\Varid{xs}\;\Varid{k}))\mathrel{\sim}\mbox{\textquotesingle}\Conid{True}{}\<[E]%
\ColumnHook
\end{hscode}\resethooks
\caption{The ``well-typed, or non-existent'' constraints.}
\label{fig:xxxOrNX}
\end{figure}

The \Edis{} counterpart of \texttt{LPUSH} and \texttt{LLEN} are therefore:\\
\begin{hscode}\SaveRestoreHook
\column{B}{@{}>{\hspre}l<{\hspost}@{}}%
\column{10}{@{}>{\hspre}l<{\hspost}@{}}%
\column{11}{@{}>{\hspre}l<{\hspost}@{}}%
\column{13}{@{}>{\hspre}l<{\hspost}@{}}%
\column{E}{@{}>{\hspre}l<{\hspost}@{}}%
\>[B]{}\Varid{lpush}\mathbin{::}{}\<[11]%
\>[11]{}(\Conid{KnownSymbol}\;\Varid{k},\Conid{Serialize}\;\Varid{a},\Conid{ListOrNX}\;\Varid{xs}\;\Varid{k})\Rightarrow {}\<[E]%
\\
\>[11]{}\Conid{Proxy}\;\Varid{k}\to \Varid{a}\to {}\<[E]%
\\
\>[11]{}\hsindent{2}{}\<[13]%
\>[13]{}\Conid{Edis}\;\Varid{xs}\;(\Conid{Set}\;\Varid{xs}\;\Varid{k}\;(\Conid{ListOf}\;\Varid{a}))\;(\Conid{R} \uplus \Conid{Integer}){}\<[E]%
\\
\>[B]{}\Varid{lpush}\;\Varid{key}\;\Varid{val}\mathrel{=}{}\<[E]%
\\
\>[B]{}\hsindent{11}{}\<[11]%
\>[11]{}\Conid{Edis}\;(\Varid{\Conid{Hedis}.lpush}\;(\Varid{encodeKey}\;\Varid{key})\;[\mskip1.5mu \Varid{encode}\;\Varid{val}\mskip1.5mu])~~,{}\<[E]%
\\[\blanklineskip]%
\>[B]{}\Varid{llen}\mathbin{::}{}\<[10]%
\>[10]{}(\Conid{KnownSymbol}\;\Varid{k},\Conid{ListOrNX}\;\Varid{xs}\;\Varid{k})\Rightarrow {}\<[E]%
\\
\>[10]{}\Conid{Proxy}\;\Varid{k}\to \Conid{Edis}\;\Varid{xs}\;\Varid{xs}\;(\Conid{R} \uplus \Conid{Integer}){}\<[E]%
\\
\>[B]{}\Varid{llen}\;\Varid{key}\mathrel{=}\Conid{Edis}\;(\Varid{\Conid{Hedis}.llen}\;(\Varid{encodeKey}\;\Varid{key}))~~.{}\<[E]%
\ColumnHook
\end{hscode}\resethooks
Similarly, the type of \ensuremath{\Varid{sadd}}, a function we have talked about a lot,
is given below:
\begin{hscode}\SaveRestoreHook
\column{B}{@{}>{\hspre}l<{\hspost}@{}}%
\column{5}{@{}>{\hspre}l<{\hspost}@{}}%
\column{10}{@{}>{\hspre}l<{\hspost}@{}}%
\column{12}{@{}>{\hspre}l<{\hspost}@{}}%
\column{E}{@{}>{\hspre}l<{\hspost}@{}}%
\>[B]{}\Varid{sadd}\mathbin{::}{}\<[10]%
\>[10]{}(\Conid{KnownSymbol}\;\Varid{k},\Conid{Serialize}\;\Varid{a},\Conid{SetOrNX}\;\Varid{xs}\;\Varid{k})\Rightarrow {}\<[E]%
\\
\>[10]{}\Conid{Proxy}\;\Varid{k}\to \Varid{a}\to {}\<[E]%
\\
\>[10]{}\hsindent{2}{}\<[12]%
\>[12]{}\Conid{Edis}\;\Varid{xs}\;(\Conid{Set}\;\Varid{xs}\;\Varid{k}\;(\Conid{SetOf}\;\Varid{a}))\;(\Conid{R} \uplus \Conid{Integer}){}\<[E]%
\\
\>[B]{}\Varid{sadd}\;\Varid{key}\;\Varid{val}\mathrel{=}{}\<[E]%
\\
\>[B]{}\hsindent{5}{}\<[5]%
\>[5]{}\Conid{Edis}\;(\Varid{\Conid{Hedis}.sadd}\;(\Varid{encodeKey}\;\Varid{key})\;[\mskip1.5mu \Varid{encode}\;\Varid{val}\mskip1.5mu])~~,{}\<[E]%
\ColumnHook
\end{hscode}\resethooks

To see a command with a more complex type, consider \ensuremath{\Varid{setnx}}, which
uses the type-level function \ensuremath{\Conid{If}} defined in Section \ref{sec:type-fun}:
\begin{hscode}\SaveRestoreHook
\column{B}{@{}>{\hspre}l<{\hspost}@{}}%
\column{5}{@{}>{\hspre}l<{\hspost}@{}}%
\column{11}{@{}>{\hspre}l<{\hspost}@{}}%
\column{12}{@{}>{\hspre}l<{\hspost}@{}}%
\column{21}{@{}>{\hspre}l<{\hspost}@{}}%
\column{E}{@{}>{\hspre}l<{\hspost}@{}}%
\>[B]{}\Varid{setnx}\mathbin{::}{}\<[11]%
\>[11]{}(\Conid{KnownSymbol}\;\Varid{k},\Conid{Serialize}\;\Varid{a})\Rightarrow \Conid{Proxy}\;\Varid{k}\to \Varid{a}\to {}\<[E]%
\\
\>[11]{}\hsindent{1}{}\<[12]%
\>[12]{}\Conid{Edis}\;\Varid{xs}\;{}\<[21]%
\>[21]{}(\Conid{If}\;(\Conid{Member}\;\Varid{xs}\;\Varid{k})\;\Varid{xs}\;(\Conid{Set}\;\Varid{xs}\;\Varid{k}\;(\Conid{StringOf}\;\Varid{a})))\;{}\<[E]%
\\
\>[21]{}(\Conid{Either}\;\Conid{Reply}\;\Conid{Bool}){}\<[E]%
\\
\>[B]{}\Varid{setnx}\;\Varid{key}\;\Varid{val}\mathrel{=}{}\<[E]%
\\
\>[B]{}\hsindent{5}{}\<[5]%
\>[5]{}\Conid{Edis}\;(\Varid{\Conid{Hedis}.setnx}\;(\Varid{encodeKey}\;\Varid{key})\;(\Varid{encode}\;\Varid{val}))~~.{}\<[E]%
\ColumnHook
\end{hscode}\resethooks
From the type one can see that \ensuremath{\Varid{setnx}\;\Varid{key}\;\Varid{val}} creates a new entry \ensuremath{(\Varid{key},\Varid{val})}
in the data store only if \ensuremath{\Varid{key}} is fresh. The type of \ensuremath{\Varid{setnx}} computes a
postcondition for static checking, as well as serving as a good documentation
for its semantics.

\subsection{Hashes}

{\em Hash} is a useful datatype supported by \Redis{}. While the \Redis{} data
store can be seen as a set of key/value pairs, a hash is itself a set of
field/value pairs. The following commands assigns a hash to key \texttt{user}.
The fields are \texttt{name}, \texttt{birthyear}, and \texttt{verified},
respectively with values \texttt{banacorn}, \texttt{1992}, and \texttt{1}.
\begin{Verbatim}[xleftmargin=.4in]
redis> hmset user name banacorn
       birthyear 1992 verified 1
OK
redis> hget user name
"banacorn"
redis> hget user birthyear
"1992"
\end{Verbatim}

For a hash to be useful, we should allow the fields to have different types. To
keep track of types of fields in a hash, \ensuremath{\Conid{HashOf}} takes a list of \ensuremath{(\Conid{Symbol},\mathbin{*})}
pairs:
\begin{hscode}\SaveRestoreHook
\column{B}{@{}>{\hspre}l<{\hspost}@{}}%
\column{E}{@{}>{\hspre}l<{\hspost}@{}}%
\>[B]{}\mathbf{data}\;\Conid{HashOf}\mathbin{::}[\mskip1.5mu (\Conid{Symbol},\mathbin{*})\mskip1.5mu]\to \mathbin{*}~~.{}\<[E]%
\ColumnHook
\end{hscode}\resethooks
By having an entry \ensuremath{(\Varid{k},\Conid{HashOF}\;\Varid{ys})} in a dictionary, we denote that the value of
key \ensuremath{\Varid{k}} is a hash whose fields and their types are specified by \ensuremath{\Varid{ys}}, which is
also a dictionary.

\begin{figure*}[t]
\begin{hscode}\SaveRestoreHook
\column{B}{@{}>{\hspre}l<{\hspost}@{}}%
\column{5}{@{}>{\hspre}l<{\hspost}@{}}%
\column{34}{@{}>{\hspre}l<{\hspost}@{}}%
\column{35}{@{}>{\hspre}l<{\hspost}@{}}%
\column{36}{@{}>{\hspre}l<{\hspost}@{}}%
\column{43}{@{}>{\hspre}l<{\hspost}@{}}%
\column{44}{@{}>{\hspre}l<{\hspost}@{}}%
\column{45}{@{}>{\hspre}l<{\hspost}@{}}%
\column{76}{@{}>{\hspre}l<{\hspost}@{}}%
\column{77}{@{}>{\hspre}l<{\hspost}@{}}%
\column{81}{@{}>{\hspre}l<{\hspost}@{}}%
\column{E}{@{}>{\hspre}l<{\hspost}@{}}%
\>[B]{}\mathbf{type}\;\Varid{family}\;\Conid{GetHash}\;(\Varid{xs}\mathbin{::}[\mskip1.5mu (\Conid{Symbol},\mathbin{*})\mskip1.5mu])\;(\Varid{k}\mathbin{::}\Conid{Symbol})\;(\Varid{f}\mathbin{::}\Conid{Symbol})\mathbin{::}\mathbin{*}\mathbf{where}{}\<[E]%
\\
\>[B]{}\hsindent{5}{}\<[5]%
\>[5]{}\Conid{GetHash}\;(\mbox{\textquotesingle}(\Varid{k},\Conid{HashOf}\;\Varid{hs}){}\<[35]%
\>[35]{}\mathrel{\,\mbox{\textquotesingle}\!\!:}\Varid{xs})\;{}\<[44]%
\>[44]{}\Varid{k}\;\Varid{f}\mathrel{=}\Conid{Get}\;\Varid{hs}\;\Varid{f}{}\<[E]%
\\
\>[B]{}\hsindent{5}{}\<[5]%
\>[5]{}\Conid{GetHash}\;(\mbox{\textquotesingle}(\Varid{l},\Varid{y}){}\<[35]%
\>[35]{}\mathrel{\,\mbox{\textquotesingle}\!\!:}\Varid{xs})\;{}\<[44]%
\>[44]{}\Varid{k}\;\Varid{f}\mathrel{=}\Conid{GetHash}\;\Varid{xs}\;\Varid{k}\;\Varid{f}{}\<[E]%
\\[\blanklineskip]%
\>[B]{}\mathbf{type}\;\Varid{family}\;\Conid{SetHash}\;(\Varid{xs}\mathbin{::}[\mskip1.5mu (\Conid{Symbol},\mathbin{*})\mskip1.5mu])\;(\Varid{k}\mathbin{::}\Conid{Symbol})\;(\Varid{f}\mathbin{::}\Conid{Symbol})\;(\Varid{a}\mathbin{::}\mathbin{*})\mathbin{::}[\mskip1.5mu (\Conid{Symbol},\mathbin{*})\mskip1.5mu]\;\mathbf{where}{}\<[E]%
\\
\>[B]{}\hsindent{5}{}\<[5]%
\>[5]{}\Conid{SetHash}\;\mbox{\textquotesingle}[~]\;{}\<[44]%
\>[44]{}\Varid{k}\;\Varid{f}\;\Varid{a}\mathrel{=}\mbox{\textquotesingle}(\Varid{k},\Conid{HashOf}\;(\Conid{Set}\;\mbox{\textquotesingle}[~]\;\Varid{f}\;\Varid{a}))\mathrel{\,\mbox{\textquotesingle}\!\!:}\mbox{\textquotesingle}[~]{}\<[E]%
\\
\>[B]{}\hsindent{5}{}\<[5]%
\>[5]{}\Conid{SetHash}\;(\mbox{\textquotesingle}(\Varid{k},\Conid{HashOf}\;\Varid{hs}){}\<[35]%
\>[35]{}\mathrel{\,\mbox{\textquotesingle}\!\!:}\Varid{xs})\;{}\<[44]%
\>[44]{}\Varid{k}\;\Varid{f}\;\Varid{a}\mathrel{=}\mbox{\textquotesingle}(\Varid{k},\Conid{HashOf}\;(\Conid{Set}\;\Varid{hs}\;{}\<[77]%
\>[77]{}\Varid{f}\;\Varid{a}))\mathrel{\,\mbox{\textquotesingle}\!\!:}\Varid{xs}{}\<[E]%
\\
\>[B]{}\hsindent{5}{}\<[5]%
\>[5]{}\Conid{SetHash}\;(\mbox{\textquotesingle}(\Varid{l},\Varid{y}){}\<[35]%
\>[35]{}\mathrel{\,\mbox{\textquotesingle}\!\!:}\Varid{xs})\;{}\<[44]%
\>[44]{}\Varid{k}\;\Varid{f}\;\Varid{a}\mathrel{=}\mbox{\textquotesingle}(\Varid{l},\Varid{y}){}\<[81]%
\>[81]{}\mathrel{\,\mbox{\textquotesingle}\!\!:}\Conid{SetHash}\;\Varid{xs}\;\Varid{k}\;\Varid{f}\;\Varid{a}{}\<[E]%
\\[\blanklineskip]%
\>[B]{}\mathbf{type}\;\Varid{family}\;\Conid{DelHash}\;(\Varid{xs}\mathbin{::}[\mskip1.5mu (\Conid{Symbol},\mathbin{*})\mskip1.5mu])\;(\Varid{k}\mathbin{::}\Conid{Symbol})\;(\Varid{f}\mathbin{::}\Conid{Symbol})\mathbin{::}[\mskip1.5mu (\Conid{Symbol},\mathbin{*})\mskip1.5mu]\;\mathbf{where}{}\<[E]%
\\
\>[B]{}\hsindent{5}{}\<[5]%
\>[5]{}\Conid{DelHash}\;\mbox{\textquotesingle}[~]\;{}\<[43]%
\>[43]{}\Varid{k}\;\Varid{f}\mathrel{=}\mbox{\textquotesingle}[~]{}\<[E]%
\\
\>[B]{}\hsindent{5}{}\<[5]%
\>[5]{}\Conid{DelHash}\;(\mbox{\textquotesingle}(\Varid{k},\Conid{HashOf}\;\Varid{hs})\mathrel{\,\mbox{\textquotesingle}\!\!:}\Varid{xs})\;{}\<[43]%
\>[43]{}\Varid{k}\;\Varid{f}\mathrel{=}\mbox{\textquotesingle}(\Varid{k},\Conid{HashOf}\;(\Conid{Del}\;\Varid{hs}\;\Varid{f}))\mathrel{\,\mbox{\textquotesingle}\!\!:}\Varid{xs}{}\<[E]%
\\
\>[B]{}\hsindent{5}{}\<[5]%
\>[5]{}\Conid{DelHash}\;(\mbox{\textquotesingle}(\Varid{l},\Varid{y}){}\<[34]%
\>[34]{}\mathrel{\,\mbox{\textquotesingle}\!\!:}\Varid{xs})\;{}\<[43]%
\>[43]{}\Varid{k}\;\Varid{f}\mathrel{=}\mbox{\textquotesingle}(\Varid{l},\Varid{y}){}\<[76]%
\>[76]{}\mathrel{\,\mbox{\textquotesingle}\!\!:}\Conid{DelHash}\;\Varid{xs}\;\Varid{k}\;\Varid{f}{}\<[E]%
\\[\blanklineskip]%
\>[B]{}\mathbf{type}\;\Varid{family}\;\Conid{MemHash}\;(\Varid{xs}\mathbin{::}[\mskip1.5mu (\Conid{Symbol},\mathbin{*})\mskip1.5mu])\;(\Varid{k}\mathbin{::}\Conid{Symbol})\;(\Varid{f}\mathbin{::}\Conid{Symbol})\mathbin{::}\Conid{Bool}\;\mathbf{where}{}\<[E]%
\\
\>[B]{}\hsindent{5}{}\<[5]%
\>[5]{}\Conid{MemHash}\;\mbox{\textquotesingle}[~]\;{}\<[45]%
\>[45]{}\Varid{k}\;\Varid{f}\mathrel{=}\mbox{\textquotesingle}\Conid{False}{}\<[E]%
\\
\>[B]{}\hsindent{5}{}\<[5]%
\>[5]{}\Conid{MemHash}\;(\mbox{\textquotesingle}(\Varid{k},\Conid{HashOf}\;\Varid{hs}){}\<[36]%
\>[36]{}\mathrel{\,\mbox{\textquotesingle}\!\!:}\Varid{xs})\;{}\<[45]%
\>[45]{}\Varid{k}\;\Varid{f}\mathrel{=}\Conid{Member}\;\Varid{hs}\;\Varid{f}{}\<[E]%
\\
\>[B]{}\hsindent{5}{}\<[5]%
\>[5]{}\Conid{MemHash}\;(\mbox{\textquotesingle}(\Varid{k},\Varid{x}){}\<[36]%
\>[36]{}\mathrel{\,\mbox{\textquotesingle}\!\!:}\Varid{xs})\;{}\<[45]%
\>[45]{}\Varid{k}\;\Varid{f}\mathrel{=}\mbox{\textquotesingle}\Conid{False}{}\<[E]%
\\
\>[B]{}\hsindent{5}{}\<[5]%
\>[5]{}\Conid{MemHash}\;(\mbox{\textquotesingle}(\Varid{l},\Varid{y}){}\<[36]%
\>[36]{}\mathrel{\,\mbox{\textquotesingle}\!\!:}\Varid{xs})\;{}\<[45]%
\>[45]{}\Varid{k}\;\Varid{f}\mathrel{=}\Conid{MemHash}\;\Varid{xs}\;\Varid{k}\;\Varid{f}{}\<[E]%
\\
\>[B]{}\hsindent{5}{}\<[5]%
\>[5]{}~~{}\<[E]%
\ColumnHook
\end{hscode}\resethooks
\vspace{-1cm}
\caption{Type-level operations for dictionaries with hashes.}
\label{fig:xxxHash}
\end{figure*}

Figure \ref{fig:xxxHash} presents some operations we need on dictionaries when
dealing with hashes. Let \ensuremath{\Varid{xs}} be a dictionary, \ensuremath{\Conid{GetHash}\;\Varid{xs}\;\Varid{k}\;\Varid{f}} returns the type
of field \ensuremath{\Varid{f}} in the hash assigned to key \ensuremath{\Varid{k}}, if both \ensuremath{\Varid{k}} and \ensuremath{\Varid{f}} exists.
\ensuremath{\Conid{SetHash}\;\Varid{xs}\;\Varid{k}\;\Varid{f}\;\Varid{a}} assigns the type \ensuremath{\Varid{a}} to the field \ensuremath{\Varid{f}} of hash \ensuremath{\Varid{k}}; if either
\ensuremath{\Varid{f}} or \ensuremath{\Varid{k}} does not exist, the hash/field is created. \ensuremath{\Conid{Del}\;\Varid{xs}\;\Varid{k}\;\Varid{f}} removes a
field, while \ensuremath{\Conid{MemHash}\;\Varid{xs}\;\Varid{k}\;\Varid{f}} checks whether the key \ensuremath{\Varid{k}} exists in \ensuremath{\Varid{xs}}, and its
value is a hash having field \ensuremath{\Varid{f}}. Their definitions make use of functions \ensuremath{\Conid{Get}},
\ensuremath{\Conid{Set}}, and \ensuremath{\Conid{Member}} defined for dictionaries.

Once those type-level functions are defined, embedding of \Hedis{} commands for
hashes is more or less routine. For example, functions \ensuremath{\Varid{hset}} and \ensuremath{\Varid{hget}}
are shown below. Note that, instead of \ensuremath{\Varid{hmset}} (available in \Hedis{}), we
provide a function \ensuremath{\Varid{hset}} that assigns fields and values one pair at a time.
\begin{hscode}\SaveRestoreHook
\column{B}{@{}>{\hspre}l<{\hspost}@{}}%
\column{3}{@{}>{\hspre}l<{\hspost}@{}}%
\column{7}{@{}>{\hspre}l<{\hspost}@{}}%
\column{10}{@{}>{\hspre}l<{\hspost}@{}}%
\column{11}{@{}>{\hspre}l<{\hspost}@{}}%
\column{12}{@{}>{\hspre}l<{\hspost}@{}}%
\column{13}{@{}>{\hspre}l<{\hspost}@{}}%
\column{E}{@{}>{\hspre}l<{\hspost}@{}}%
\>[B]{}\Varid{hset}{}\<[7]%
\>[7]{}\mathbin{::}(\Conid{KnownSymbol}\;\Varid{k},\Conid{KnownSymbol}\;\Varid{f},{}\<[E]%
\\
\>[7]{}\hsindent{4}{}\<[11]%
\>[11]{}\Conid{Serialize}\;\Varid{a},\Conid{HashOrNX}\;\Varid{xs}\;\Varid{k}){}\<[E]%
\\
\>[7]{}\Rightarrow \Conid{Proxy}\;\Varid{k}\to \Conid{Proxy}\;\Varid{f}\to \Varid{a}{}\<[E]%
\\
\>[7]{}\to \Conid{Edis}\;\Varid{xs}\;(\Conid{SetHash}\;\Varid{xs}\;\Varid{k}\;\Varid{f}\;(\Conid{StringOf}\;\Varid{a}))\;(\Conid{R} \uplus \Conid{Bool}){}\<[E]%
\\
\>[B]{}\Varid{hset}\;\Varid{key}\;\Varid{field}\;\Varid{val}\mathrel{=}{}\<[E]%
\\
\>[B]{}\hsindent{3}{}\<[3]%
\>[3]{}\Conid{Edis}\;({}\<[11]%
\>[11]{}\Varid{\Conid{Hedis}.hset}\;(\Varid{encodeKey}\;\Varid{key})\;{}\<[E]%
\\
\>[11]{}\hsindent{1}{}\<[12]%
\>[12]{}(\Varid{encodeKey}\;\Varid{field})\;(\Varid{encode}\;\Varid{val}))~~,{}\<[E]%
\\[\blanklineskip]%
\>[B]{}\Varid{hget}\mathbin{::}{}\<[10]%
\>[10]{}({}\<[13]%
\>[13]{}\Conid{KnownSymbol}\;\Varid{k},\Conid{KnownSymbol}\;\Varid{f},\Conid{Serialize}\;\Varid{a},{}\<[E]%
\\
\>[13]{}\Conid{StringOf}\;\Varid{a}\mathrel{\sim}\Conid{GetHash}\;\Varid{xs}\;\Varid{k}\;\Varid{f})\Rightarrow {}\<[E]%
\\
\>[10]{}\Conid{Proxy}\;\Varid{k}\to \Conid{Proxy}\;\Varid{f}\to \Conid{Edis}\;\Varid{xs}\;\Varid{xs}\;(\Conid{R} \uplus \Conid{Maybe}\;\Varid{a}){}\<[E]%
\\
\>[B]{}\Varid{hget}\;\Varid{key}\;\Varid{field}\mathrel{=}{}\<[E]%
\\
\>[B]{}\hsindent{3}{}\<[3]%
\>[3]{}\Conid{Edis}\;({}\<[11]%
\>[11]{}\Varid{\Conid{Hedis}.hget}\;(\Varid{encodeKey}\;\Varid{key})\;(\Varid{encodeKey}\;\Varid{field})\mathbin{>\!\!>\!\!=}{}\<[E]%
\\
\>[11]{}\Varid{decodeAsMaybe})~~,{}\<[E]%
\ColumnHook
\end{hscode}\resethooks
where
\begin{hscode}\SaveRestoreHook
\column{B}{@{}>{\hspre}l<{\hspost}@{}}%
\column{3}{@{}>{\hspre}l<{\hspost}@{}}%
\column{E}{@{}>{\hspre}l<{\hspost}@{}}%
\>[B]{}\Varid{decodeAsMaybe}\mathbin{::}\Conid{Serialize}\;\Varid{a}\Rightarrow (\Conid{R} \uplus \Conid{Maybe}\;\Conid{ByteString})\to {}\<[E]%
\\
\>[B]{}\hsindent{3}{}\<[3]%
\>[3]{}\Conid{Redis}\;(\Conid{R} \uplus \Conid{Maybe}\;\Varid{a})~~,{}\<[E]%
\ColumnHook
\end{hscode}\resethooks
using the function \ensuremath{\Varid{decode}}
mentioned in Section \ref{sec:proxy-key}, parses the \ensuremath{\Conid{ByteString}} in
\ensuremath{\Conid{R} \uplus \Conid{Maybe}\;\anonymous } to type \ensuremath{\Varid{a}}. The definition is a bit tedious but
routine.

\hide{Note that, instead of \ensuremath{\Varid{hmset}} (available in \Hedis{}), we
provide a function \ensuremath{\Varid{hset}} that assigns fields and values one pair at at time.}
We will talk about difficulties of implementing \ensuremath{\Varid{hmset}} in
Section~\ref{sec:discussions}.

\subsection{Assertions}
\label{sec:assertions}

%

Finally, the creation/update behavior of \Redis{} functions is, in our opinion,
very error-prone. It might be preferable if we can explicit declare some new
keys, after ensuring that they do not already exist (in our types). This can be done below:
\begin{hscode}\SaveRestoreHook
\column{B}{@{}>{\hspre}l<{\hspost}@{}}%
\column{13}{@{}>{\hspre}l<{\hspost}@{}}%
\column{E}{@{}>{\hspre}l<{\hspost}@{}}%
\>[B]{}\Varid{declare}\mathbin{::}{}\<[13]%
\>[13]{}(\Conid{KnownSymbol}\;\Varid{k},\Conid{Member}\;\Varid{xs}\;\Varid{k}\mathrel{\sim}\Conid{False})\Rightarrow {}\<[E]%
\\
\>[13]{}\Conid{Proxy}\;\Varid{k}\to \Conid{Proxy}\;\Varid{a}\to \Conid{Edis}\;\Varid{xs}\;(\Conid{Set}\;\Varid{xs}\;\Varid{k}\;\Varid{a})\;(){}\<[E]%
\\
\>[B]{}\Varid{declare}\;\Varid{key}\;\Varid{typ}\mathrel{=}\Conid{Edis}\;(\Varid{return}\;())~~.{}\<[E]%
\ColumnHook
\end{hscode}\resethooks
The command \ensuremath{\Varid{declare}\;\Varid{key}\;\Varid{typ}}, where \ensuremath{\Varid{typ}} is the proxy of \ensuremath{\Varid{a}}, adds a fresh
 \ensuremath{\Varid{key}} with type \ensuremath{\Varid{a}} into the dictionary. Notice that \ensuremath{\Varid{declare}} does nothing at
 term level, but simply returns \ensuremath{()}, since it only has effects on types.


Another command for type level assertion, \ensuremath{\Varid{start}}, initializes the dictionary to
 the empty list, comes in handy when starting a series of \Edis{} commands:

\begin{hscode}\SaveRestoreHook
\column{B}{@{}>{\hspre}l<{\hspost}@{}}%
\column{E}{@{}>{\hspre}l<{\hspost}@{}}%
\>[B]{}\Varid{start}\mathbin{::}\Conid{Edis}\;\mbox{\textquotesingle}[~]\;\mbox{\textquotesingle}[~]\;(){}\<[E]%
\\
\>[B]{}\Varid{start}\mathrel{=}\Conid{Edis}\;(\Varid{return}\;())~~.{}\<[E]%
\ColumnHook
\end{hscode}\resethooks

\subsection{A Slightly Larger Example}

We present a slightly larger example as a summary. The task is to store a queue of
messages in \Redis{}. Messages are represented by a \ensuremath{\Conid{ByteString}} and an
\ensuremath{\Conid{Integer}} identifier:%
\footnote{\ensuremath{\Conid{Message}} is made an instance of \ensuremath{\Conid{Generic}} in order to use the
generic implementation of methods of \ensuremath{\Conid{Serialize}}.}
\begin{hscode}\SaveRestoreHook
\column{B}{@{}>{\hspre}l<{\hspost}@{}}%
\column{20}{@{}>{\hspre}l<{\hspost}@{}}%
\column{35}{@{}>{\hspre}l<{\hspost}@{}}%
\column{E}{@{}>{\hspre}l<{\hspost}@{}}%
\>[B]{}\mathbf{data}\;\Conid{Message}\mathrel{=}\Conid{Msg}\;\Conid{ByteString}\;\Conid{Integer}{}\<[E]%
\\
\>[B]{}\hsindent{20}{}\<[20]%
\>[20]{}\mathbf{deriving}\;(\Conid{Show},\Conid{Generic})~~,{}\<[E]%
\\
\>[B]{}\mathbf{instance}\;\Conid{Serialize}\;\Conid{Message}\;\mathbf{where}\;{}\<[35]%
\>[35]{}~~.{}\<[E]%
\ColumnHook
\end{hscode}\resethooks

In the data store, the queue is represented by a list. Before pushing a message
into the queue, we increment \ensuremath{\Varid{counter}}, a key storing a counter, and use it as the
identifier of the message:
\begin{hscode}\SaveRestoreHook
\column{B}{@{}>{\hspre}l<{\hspost}@{}}%
\column{10}{@{}>{\hspre}l<{\hspost}@{}}%
\column{11}{@{}>{\hspre}l<{\hspost}@{}}%
\column{13}{@{}>{\hspre}l<{\hspost}@{}}%
\column{14}{@{}>{\hspre}l<{\hspost}@{}}%
\column{E}{@{}>{\hspre}l<{\hspost}@{}}%
\>[B]{}\Varid{push}\mathbin{::}{}\<[10]%
\>[10]{}({}\<[14]%
\>[14]{}\Conid{StringOfIntegerOrNX}\;\Varid{xs}\;\text{\tt \char34 counter\char34},{}\<[E]%
\\
\>[14]{}\Conid{ListOrNX}\;\Varid{xs}\;\text{\tt \char34 queue\char34})\Rightarrow {}\<[E]%
\\
\>[10]{}\Conid{ByteString}\to \Conid{Edis}\;\Varid{xs}\;(\Conid{Set}\;\Varid{xs}\;\text{\tt \char34 queue\char34}{}\<[E]%
\\
\>[10]{}\hsindent{1}{}\<[11]%
\>[11]{}(\Conid{ListOf}\;\Conid{Message}))\;(\Conid{R} \uplus \Conid{Integer}){}\<[E]%
\\
\>[B]{}\Varid{push}\;\Varid{msg}\mathrel{=}{}\<[13]%
\>[13]{}\Varid{incr}\;\Varid{kCounter}\mathbin{`\Varid{bind}`}\lambda \Varid{i}\to {}\<[E]%
\\
\>[13]{}\Varid{lpush}\;\Varid{kQueue}\;(\Conid{Msg}\;\Varid{msg}\;(\Varid{fromRight}\;\Varid{i}))~~,{}\<[E]%
\ColumnHook
\end{hscode}\resethooks
where \ensuremath{\Varid{fromRight}\mathbin{::}\Conid{Either}\;\Varid{a}\;\Varid{b}\to \Varid{b}} extracts the value wrapped by constructor
\ensuremath{\Conid{Right}}, and the constraint \ensuremath{\Conid{StringOfIntegerOrNX}} \ensuremath{\Varid{xs}\;\Varid{k}} holds if either \ensuremath{\Varid{k}}
appears in \ensuremath{\Varid{xs}} and is converted from an \ensuremath{\Conid{Integer}}, or \ensuremath{\Varid{k}} does not
appear in \ensuremath{\Varid{xs}}. For brevity, the proxies are given names: \\
\begin{hscode}\SaveRestoreHook
\column{B}{@{}>{\hspre}l<{\hspost}@{}}%
\column{14}{@{}>{\hspre}l<{\hspost}@{}}%
\column{E}{@{}>{\hspre}l<{\hspost}@{}}%
\>[B]{}\Varid{kCounter}\mathbin{::}{}\<[14]%
\>[14]{}\Conid{Proxy}\;\text{\tt \char34 counter\char34}{}\<[E]%
\\
\>[B]{}\Varid{kCounter}\mathrel{=}{}\<[14]%
\>[14]{}\Conid{Proxy}~~,{}\<[E]%
\ColumnHook
\end{hscode}\resethooks
\begin{hscode}\SaveRestoreHook
\column{B}{@{}>{\hspre}l<{\hspost}@{}}%
\column{12}{@{}>{\hspre}l<{\hspost}@{}}%
\column{E}{@{}>{\hspre}l<{\hspost}@{}}%
\>[B]{}\Varid{kQueue}\mathbin{::}{}\<[12]%
\>[12]{}\Conid{Proxy}\;\text{\tt \char34 queue\char34}{}\<[E]%
\\
\>[B]{}\Varid{kQueue}\mathrel{=}{}\<[12]%
\>[12]{}\Conid{Proxy}~~.{}\<[E]%
\ColumnHook
\end{hscode}\resethooks
To pop a message we use the function \ensuremath{\Varid{rpop}} which, given a key associated with
a list, extracts the rightmost element of the list
\begin{hscode}\SaveRestoreHook
\column{B}{@{}>{\hspre}l<{\hspost}@{}}%
\column{9}{@{}>{\hspre}l<{\hspost}@{}}%
\column{10}{@{}>{\hspre}l<{\hspost}@{}}%
\column{20}{@{}>{\hspre}l<{\hspost}@{}}%
\column{E}{@{}>{\hspre}l<{\hspost}@{}}%
\>[B]{}\Varid{pop}\mathbin{::}{}\<[9]%
\>[9]{}(\Conid{Get}\;\Varid{xs}\;\text{\tt \char34 queue\char34}\mathrel{\sim}\Conid{ListOf}\;\Conid{Message})\Rightarrow {}\<[E]%
\\
\>[9]{}\Conid{Edis}\;\Varid{xs}\;\Varid{xs}\;(\Conid{R} \uplus \Conid{Maybe}\;\Conid{Message}){}\<[E]%
\\
\>[B]{}\Varid{pop}\mathrel{=}\Varid{rpop}\;\Varid{kQueue}~~,{}\<[E]%
\\[\blanklineskip]%
\>[B]{}\Varid{rpop}\mathbin{::}{}\<[10]%
\>[10]{}(\Conid{KnownSymbol}\;\Varid{k},\Conid{Serialize}\;\Varid{a},\Conid{Get}\;\Varid{xs}\;\Varid{k}\mathrel{\sim}\Conid{ListOf}\;\Varid{a})\Rightarrow {}\<[E]%
\\
\>[10]{}\Conid{Proxy}\;\Varid{k}\to \Conid{Edis}\;\Varid{xs}\;\Varid{xs}\;(\Conid{R} \uplus \Conid{Maybe}\;\Varid{a}){}\<[E]%
\\
\>[B]{}\Varid{rpop}\;\Varid{key}\mathrel{=}\Conid{Edis}\;({}\<[20]%
\>[20]{}\Varid{\Conid{Hedis}.rpop}\;(\Varid{encodeKey}\;\Varid{key})\mathbin{>\!\!>\!\!=}{}\<[E]%
\\
\>[20]{}\Varid{decodeAsMaybe})~~.{}\<[E]%
\ColumnHook
\end{hscode}\resethooks
Our sample program is shown below:
\begin{hscode}\SaveRestoreHook
\column{B}{@{}>{\hspre}l<{\hspost}@{}}%
\column{4}{@{}>{\hspre}c<{\hspost}@{}}%
\column{4E}{@{}l@{}}%
\column{9}{@{}>{\hspre}l<{\hspost}@{}}%
\column{27}{@{}>{\hspre}l<{\hspost}@{}}%
\column{E}{@{}>{\hspre}l<{\hspost}@{}}%
\>[B]{}\Varid{prog}\mathrel{=}{}\<[9]%
\>[9]{}\Varid{declare}\;\Varid{kCounter}\;{}\<[27]%
\>[27]{}(\Conid{Proxy}\mathbin{::}\Conid{Proxy}\;\Conid{Integer}){}\<[E]%
\\
\>[B]{}\hsindent{4}{}\<[4]%
\>[4]{}\mathbin{>\!\!>\!\!>}{}\<[4E]%
\>[9]{}\Varid{declare}\;\Varid{kQueue}\;{}\<[27]%
\>[27]{}(\Conid{Proxy}\mathbin{::}\Conid{Proxy}\;(\Conid{ListOf}\;\Conid{Message})){}\<[E]%
\\
\>[B]{}\hsindent{4}{}\<[4]%
\>[4]{}\mathbin{>\!\!>\!\!>}{}\<[4E]%
\>[9]{}\Varid{push}\;\text{\tt \char34 hello\char34}{}\<[E]%
\\
\>[B]{}\hsindent{4}{}\<[4]%
\>[4]{}\mathbin{>\!\!>\!\!>}{}\<[4E]%
\>[9]{}\Varid{push}\;\text{\tt \char34 world\char34}{}\<[E]%
\\
\>[B]{}\hsindent{4}{}\<[4]%
\>[4]{}\mathbin{>\!\!>\!\!>}{}\<[4E]%
\>[9]{}\Varid{pop}~~,{}\<[E]%
\ColumnHook
\end{hscode}\resethooks
where the monadic sequencing operator \ensuremath{(\mathbin{>\!\!>\!\!>})} is defined by:
\begin{hscode}\SaveRestoreHook
\column{B}{@{}>{\hspre}l<{\hspost}@{}}%
\column{E}{@{}>{\hspre}l<{\hspost}@{}}%
\>[B]{}(\mathbin{>\!\!>\!\!>})\mathbin{::}\Conid{IMonad}\;\Varid{m}\Rightarrow \Varid{m}\;\Varid{p}\;\Varid{q}\;\Varid{a}\to \Varid{m}\;\Varid{q}\;\Varid{r}\;\Varid{b}\to \Varid{m}\;\Varid{p}\;\Varid{r}\;\Varid{b}{}\<[E]%
\\
\>[B]{}\Varid{m}_{1}\mathbin{>\!\!>\!\!>}\Varid{m}_{2}\mathrel{=}\Varid{m}_{1}\mathbin{`\Varid{bind}`}(\lambda \anonymous \to \Varid{m}_{2})~~.{}\<[E]%
\ColumnHook
\end{hscode}\resethooks
Use of \ensuremath{\Varid{declare}} in \ensuremath{\Varid{prog}} ensures that neither \ensuremath{\text{\tt \char34 counter\char34}} nor \ensuremath{\text{\tt \char34 queue\char34}} exist
before the execution of \ensuremath{\Varid{prog}}. The program simply stores two strings in \ensuremath{\text{\tt \char34 queue\char34}}, before extracting the first string. GHC is able to infer the type of \ensuremath{\Varid{prog}}:
\begin{hscode}\SaveRestoreHook
\column{B}{@{}>{\hspre}l<{\hspost}@{}}%
\column{15}{@{}>{\hspre}l<{\hspost}@{}}%
\column{22}{@{}>{\hspre}l<{\hspost}@{}}%
\column{27}{@{}>{\hspre}l<{\hspost}@{}}%
\column{E}{@{}>{\hspre}l<{\hspost}@{}}%
\>[B]{}\Varid{prog}\mathbin{::}\Conid{Edis}\;{}\<[15]%
\>[15]{}\mbox{\textquotesingle}[~]\;\mbox{\textquotesingle}[{}\<[27]%
\>[27]{}\mbox{\textquotesingle}(\text{\tt \char34 counter\char34},\Conid{Integer}),{}\<[E]%
\\
\>[15]{}\hsindent{7}{}\<[22]%
\>[22]{}\quad\!\mbox{\textquotesingle}(\text{\tt \char34 queue\char34},\Conid{ListOf}\;\Conid{Message})]\;{}\<[E]%
\\
\>[15]{}(\Conid{R} \uplus \Conid{Maybe}\;\Conid{Message})~~.{}\<[E]%
\ColumnHook
\end{hscode}\resethooks

To get things going, the main program builds a connection with the \Redis{}
server, runs \ensuremath{\Varid{prog}}, and prints the result:
\begin{hscode}\SaveRestoreHook
\column{B}{@{}>{\hspre}l<{\hspost}@{}}%
\column{12}{@{}>{\hspre}l<{\hspost}@{}}%
\column{20}{@{}>{\hspre}l<{\hspost}@{}}%
\column{E}{@{}>{\hspre}l<{\hspost}@{}}%
\>[B]{}\Varid{main}\mathbin{::}\Conid{IO}\;(){}\<[E]%
\\
\>[B]{}\Varid{main}\mathrel{=}\mathbf{do}\;{}\<[12]%
\>[12]{}\Varid{conn}{}\<[20]%
\>[20]{}\leftarrow \Varid{connect}\;\Varid{defaultConnectInfo}{}\<[E]%
\\
\>[12]{}\Varid{result}{}\<[20]%
\>[20]{}\leftarrow \Varid{runRedis}\;\Varid{conn}\;(\Varid{unEdis}\;(\Varid{start}\mathbin{>\!\!>\!\!>}\Varid{prog})){}\<[E]%
\\
\>[12]{}\Varid{print}\;\Varid{result}~~.{}\<[E]%
\ColumnHook
\end{hscode}\resethooks
The command \ensuremath{\Varid{start}} in \ensuremath{\Varid{main}} guarantees that the
program is given a fresh run without previously defined keys at all.
All type-level constraints in \ensuremath{\Varid{start}\mathbin{>\!\!>\!\!>}\Varid{prog}} are stripped away
by \ensuremath{\Varid{unEdis}}. The untyped program stored in \ensuremath{\Conid{Edis}}, of type \ensuremath{\Conid{Redis}\;(\Conid{R} \uplus \Conid{Maybe}\;\Conid{Message})}, is passed to the \Redis{} function
\ensuremath{\Varid{runRedis}}, of type \ensuremath{\Conid{Connection}\to \Conid{Redis}\;\Varid{a}\to \Conid{IO}\;\Varid{a}}. In this case the output
is \ensuremath{\Conid{Right}\;(\Conid{Just}\;\text{\tt \char34 hello\char34})}.




%% file: sections/Discussions.tex
%
%
\makeatletter
\@ifundefined{lhs2tex.lhs2tex.sty.read}%
  {\@namedef{lhs2tex.lhs2tex.sty.read}{}%
   \newcommand\SkipToFmtEnd{}%
   \newcommand\EndFmtInput{}%
   \long\def\SkipToFmtEnd#1\EndFmtInput{}%
  }\SkipToFmtEnd

\newcommand\ReadOnlyOnce[1]{\@ifundefined{#1}{\@namedef{#1}{}}\SkipToFmtEnd}
\usepackage{amstext}
\usepackage{amssymb}
\usepackage{stmaryrd}
\DeclareFontFamily{OT1}{cmtex}{}
\DeclareFontShape{OT1}{cmtex}{m}{n}
  {<5><6><7><8>cmtex8
   <9>cmtex9
   <10><10.95><12><14.4><17.28><20.74><24.88>cmtex10}{}
\DeclareFontShape{OT1}{cmtex}{m}{it}
  {<-> ssub * cmtt/m/it}{}
\newcommand{\texfamily}{\fontfamily{cmtex}\selectfont}
\DeclareFontShape{OT1}{cmtt}{bx}{n}
  {<5><6><7><8>cmtt8
   <9>cmbtt9
   <10><10.95><12><14.4><17.28><20.74><24.88>cmbtt10}{}
\DeclareFontShape{OT1}{cmtex}{bx}{n}
  {<-> ssub * cmtt/bx/n}{}
\newcommand{\tex}[1]{\text{\texfamily#1}}	

\newcommand{\Sp}{\hskip.33334em\relax}

\newcommand{\Conid}[1]{\mathit{#1}}
\newcommand{\Varid}[1]{\mathit{#1}}
\newcommand{\anonymous}{\kern0.06em \vbox{\hrule\@width.5em}}
\newcommand{\plus}{\mathbin{+\!\!\!+}}
\newcommand{\bind}{\mathbin{>\!\!\!>\mkern-6.7mu=}}
\newcommand{\rbind}{\mathbin{=\mkern-6.7mu<\!\!\!<}}
\newcommand{\sequ}{\mathbin{>\!\!\!>}}
\renewcommand{\leq}{\leqslant}
\renewcommand{\geq}{\geqslant}
\usepackage{polytable}

\@ifundefined{mathindent}%
  {\newdimen\mathindent\mathindent\leftmargini}%
  {}%

\def\resethooks{%
  \global\let\SaveRestoreHook\empty
  \global\let\ColumnHook\empty}
\newcommand*{\savecolumns}[1][default]%
  {\g@addto@macro\SaveRestoreHook{\savecolumns[#1]}}
\newcommand*{\restorecolumns}[1][default]%
  {\g@addto@macro\SaveRestoreHook{\restorecolumns[#1]}}
\newcommand*{\aligncolumn}[2]%
  {\g@addto@macro\ColumnHook{\column{#1}{#2}}}

\resethooks

\newcommand{\onelinecommentchars}{\quad-{}- }
\newcommand{\commentbeginchars}{\enskip\{-}
\newcommand{\commentendchars}{-\}\enskip}

\newcommand{\visiblecomments}{%
  \let\onelinecomment=\onelinecommentchars
  \let\commentbegin=\commentbeginchars
  \let\commentend=\commentendchars}

\newcommand{\invisiblecomments}{%
  \let\onelinecomment=\empty
  \let\commentbegin=\empty
  \let\commentend=\empty}

\visiblecomments

\newlength{\blanklineskip}
\setlength{\blanklineskip}{0.66084ex}

\newcommand{\hsindent}[1]{\quad}
\let\hspre\empty
\let\hspost\empty
\newcommand{\NB}{\textbf{NB}}
\newcommand{\Todo}[1]{$\langle$\textbf{To do:}~#1$\rangle$}

\EndFmtInput
\makeatother
%
%
%
%
%
%
%
%
%
\ReadOnlyOnce{polycode.fmt}%
\makeatletter

\newcommand{\hsnewpar}[1]%
  {{\parskip=0pt\parindent=0pt\par\vskip #1\noindent}}

\newcommand{\hscodestyle}{}


\newcommand{\sethscode}[1]%
  {\expandafter\let\expandafter\hscode\csname #1\endcsname
   \expandafter\let\expandafter\endhscode\csname end#1\endcsname}


\newenvironment{compathscode}%
  {\par\noindent
   \advance\leftskip\mathindent
   \hscodestyle
   \let\\=\@normalcr
   \let\hspre\(\let\hspost\)%
   \pboxed}%
  {\endpboxed\)%
   \par\noindent
   \ignorespacesafterend}

\newcommand{\compaths}{\sethscode{compathscode}}


\newenvironment{plainhscode}%
  {\hsnewpar\abovedisplayskip
   \advance\leftskip\mathindent
   \hscodestyle
   \let\hspre\(\let\hspost\)%
   \pboxed}%
  {\endpboxed%
   \hsnewpar\belowdisplayskip
   \ignorespacesafterend}

\newenvironment{oldplainhscode}%
  {\hsnewpar\abovedisplayskip
   \advance\leftskip\mathindent
   \hscodestyle
   \let\\=\@normalcr
   \(\pboxed}%
  {\endpboxed\)%
   \hsnewpar\belowdisplayskip
   \ignorespacesafterend}


\newcommand{\plainhs}{\sethscode{plainhscode}}
\newcommand{\oldplainhs}{\sethscode{oldplainhscode}}
\plainhs


\newenvironment{arrayhscode}%
  {\hsnewpar\abovedisplayskip
   \advance\leftskip\mathindent
   \hscodestyle
   \let\\=\@normalcr
   \(\parray}%
  {\endparray\)%
   \hsnewpar\belowdisplayskip
   \ignorespacesafterend}

\newcommand{\arrayhs}{\sethscode{arrayhscode}}


\newenvironment{mathhscode}%
  {\parray}{\endparray}

\newcommand{\mathhs}{\sethscode{mathhscode}}


\newenvironment{texthscode}%
  {\(\parray}{\endparray\)}

\newcommand{\texths}{\sethscode{texthscode}}


\def\codeframewidth{\arrayrulewidth}
\RequirePackage{calc}

\newenvironment{framedhscode}%
  {\parskip=\abovedisplayskip\par\noindent
   \hscodestyle
   \arrayrulewidth=\codeframewidth
   \tabular{@{}|p{\linewidth-2\arraycolsep-2\arrayrulewidth-2pt}|@{}}%
   \hline\framedhslinecorrect\\{-1.5ex}%
   \let\endoflinesave=\\
   \let\\=\@normalcr
   \(\pboxed}%
  {\endpboxed\)%
   \framedhslinecorrect\endoflinesave{.5ex}\hline
   \endtabular
   \parskip=\belowdisplayskip\par\noindent
   \ignorespacesafterend}

\newcommand{\framedhslinecorrect}[2]%
  {#1[#2]}

\newcommand{\framedhs}{\sethscode{framedhscode}}


\newenvironment{inlinehscode}%
  {\(\def\column##1##2{}%
   \let\>\undefined\let\<\undefined\let\\\undefined
   \newcommand\>[1][]{}\newcommand\<[1][]{}\newcommand\\[1][]{}%
   \def\fromto##1##2##3{##3}%
   \def\nextline{}}{\) }%

\newcommand{\inlinehs}{\sethscode{inlinehscode}}


\newenvironment{joincode}%
  {\let\orighscode=\hscode
   \let\origendhscode=\endhscode
   \def\endhscode{\def\hscode{\endgroup\def\@currenvir{hscode}\\}\begingroup}
   \orighscode\def\hscode{\endgroup\def\@currenvir{hscode}}}%
  {\origendhscode
   \global\let\hscode=\orighscode
   \global\let\endhscode=\origendhscode}%

\makeatother
\EndFmtInput

\ReadOnlyOnce{Formatting.fmt}%
\makeatletter

\let\Varid\mathit
\let\Conid\mathsf

\def\commentbegin{\quad\{\ }
\def\commentend{\}}

\newcommand{\ty}[1]{\Conid{#1}}
\newcommand{\con}[1]{\Conid{#1}}
\newcommand{\id}[1]{\Varid{#1}}
\newcommand{\cl}[1]{\Varid{#1}}
\newcommand{\opsym}[1]{\mathrel{#1}}

\newcommand\Keyword[1]{\textbf{\textsf{#1}}}
\newcommand\Hide{\mathbin{\downarrow}}
\newcommand\Reveal{\mathbin{\uparrow}}


\makeatother
\EndFmtInput

\section{Discussion}
\label{sec:discussions}


\paragraph{Returning Inferrable Types.} \texttt{GET} is yet another command that
is invokable only under a ``well-typed or non-existent'' precondition,
mentioned in Section~\ref{sec:disjunctive-constraints}. It fetches the value of
a key and, if the key does not exist, returns a special value \texttt{nil}. An
error is raised if the value is not a string. In \Edis{} the situation is made
slightly complex, since we parse the string to the type it was supposed to have
encoded from. The \Edis{} version of \ensuremath{\Varid{get}} could be typed:
\begin{hscode}\SaveRestoreHook
\column{B}{@{}>{\hspre}l<{\hspost}@{}}%
\column{9}{@{}>{\hspre}l<{\hspost}@{}}%
\column{E}{@{}>{\hspre}l<{\hspost}@{}}%
\>[B]{}\Varid{get}\mathbin{::}{}\<[9]%
\>[9]{}(\Conid{KnownSymbol}\;\Varid{k},\Conid{Serialize}\;\Varid{a},\Conid{StringOrNX}\;\Varid{xs}\;\Varid{k})\Rightarrow {}\<[E]%
\\
\>[9]{}\Conid{Proxy}\;\Varid{k}\to \Conid{Edis}\;\Varid{xs}\;\Varid{xs}\;(\Conid{R} \uplus \Conid{Maybe}\;\Varid{a})~~.{}\<[E]%
\ColumnHook
\end{hscode}\resethooks
where \ensuremath{\Conid{StringOrNX}} is defined in Figure~\ref{fig:xxxOrNX}.

The problem with such typing, however, is that \ensuremath{\Varid{a}} cannot be inferred from \ensuremath{\Varid{xs}}
and \ensuremath{\Varid{k}} when \ensuremath{\Varid{k}} does not appear in \ensuremath{\Varid{xs}}. In such situations, to avoid Haskell
complaining about ambiguous type, \ensuremath{\Varid{a}} has to be specified by the caller of
\ensuremath{\Varid{get}}. The user will then be forced to spell out the complete type signature,
only to make \ensuremath{\Varid{a}} explicit.

We think it is more reasonable to enforce that, when \ensuremath{\Varid{get}} is called, the key
should exist in the data store. Thus \ensuremath{\Varid{get}} in \Redis{} has the following type:
\begin{hscode}\SaveRestoreHook
\column{B}{@{}>{\hspre}l<{\hspost}@{}}%
\column{9}{@{}>{\hspre}l<{\hspost}@{}}%
\column{E}{@{}>{\hspre}l<{\hspost}@{}}%
\>[B]{}\Varid{get}\mathbin{::}{}\<[9]%
\>[9]{}(\Conid{KnownSymbol}\;\Varid{k},\Conid{Serialize}\;\Varid{a},\Conid{Get}\;\Varid{xs}\;\Varid{k}\mathrel{\sim}\Conid{StringOf}\;\Varid{a})\Rightarrow {}\<[E]%
\\
\>[9]{}\Conid{Proxy}\;\Varid{k}\to \Conid{Edis}\;\Varid{xs}\;\Varid{xs}\;(\Conid{R} \uplus \Conid{Maybe}\;\Varid{a})~~,{}\<[E]%
\ColumnHook
\end{hscode}\resethooks
which requires that \ensuremath{(\Varid{k},\Varid{a})} presents in \ensuremath{\Varid{xs}} and thus \ensuremath{\Varid{a}} is inferrable from
\ensuremath{\Varid{xs}} and \ensuremath{\Varid{k}}.

\paragraph{Variable Number of Input/Outputs.} Recall that, in
Section~\ref{sec:proxy-key}, the \Redis{} command \texttt{DEL} takes a variable
number of keys, while our \Edis{} counterpart takes only one. Some \Redis{}
commands take a variable number of arguments as inputs, and some returns
multiple results. Most of them are accurately implemented in \Hedis{}. For
another example of a variable-number input command, the type of \ensuremath{\Varid{sinter}} in
\Hedis{} is shown below:
\begin{hscode}\SaveRestoreHook
\column{B}{@{}>{\hspre}l<{\hspost}@{}}%
\column{E}{@{}>{\hspre}l<{\hspost}@{}}%
\>[B]{}\Varid{\Conid{Hedis}.sinter}\mathbin{::}[\mskip1.5mu \Conid{ByteString}\mskip1.5mu]\to \Conid{Redis}\;(\Conid{R} \uplus [\mskip1.5mu \Conid{ByteString}\mskip1.5mu])~~.{}\<[E]%
\ColumnHook
\end{hscode}\resethooks
It takes a list of keys, values of which are all supposed to be sets, and
computes their intersection (the returned list is the intersected set).

In \Edis{}, for a function to accept a list of keys as input, we have to
specify that all the keys are in the class \ensuremath{\Conid{KnownSymbol}}. It can be done by
defining a datatype, indexed by the keys, serving as a witness that they
are all in \ensuremath{\Conid{KnownSymbol}}. We currently have not implemented such feature and
leave it as a possible future work. For now, we offer commands that take fixed
numbers of inputs. The \Edis{} version of \ensuremath{\Varid{sinter}} has type:
\begin{hscode}\SaveRestoreHook
\column{B}{@{}>{\hspre}l<{\hspost}@{}}%
\column{12}{@{}>{\hspre}l<{\hspost}@{}}%
\column{15}{@{}>{\hspre}l<{\hspost}@{}}%
\column{E}{@{}>{\hspre}l<{\hspost}@{}}%
\>[B]{}\Varid{sinter}\mathbin{::}{}\<[12]%
\>[12]{}({}\<[15]%
\>[15]{}\Conid{KnownSymbol}\;\Varid{k}_{1},\Conid{KnownSymbol}\;\Varid{k}_{2},\Conid{Serialize}\;\Varid{a},{}\<[E]%
\\
\>[15]{}\Conid{SetOf}\;\Varid{x}\mathrel{\sim}\Conid{Get}\;\Varid{xs}\;\Varid{k}_{1},\Conid{SetOf}\;\Varid{x}\mathrel{\sim}\Conid{Get}\;\Varid{xs}\;\Varid{k}_{2})\Rightarrow {}\<[E]%
\\
\>[12]{}\Conid{Proxy}\;\Varid{k}_{1}\to \Conid{Proxy}\;\Varid{k}_{2}\to \Conid{Edis}\;\Varid{xs}\;\Varid{xs}\;(\Conid{R} \uplus [\mskip1.5mu \Varid{a}\mskip1.5mu])~~.{}\<[E]%
\ColumnHook
\end{hscode}\resethooks

The function \ensuremath{\Varid{hmset}} in \Hedis{} allows one to set the values of many fields
in a hash, while \ensuremath{\Varid{hgetall}} returns all the field-value pairs of a hash. They
have the following types:
\begin{hscode}\SaveRestoreHook
\column{B}{@{}>{\hspre}l<{\hspost}@{}}%
\column{4}{@{}>{\hspre}l<{\hspost}@{}}%
\column{66}{@{}>{\hspre}l<{\hspost}@{}}%
\column{E}{@{}>{\hspre}l<{\hspost}@{}}%
\>[B]{}\Varid{\Conid{Hedis}.hmset}\mathbin{::}\Conid{ByteString}\to {}\<[E]%
\\
\>[B]{}\hsindent{4}{}\<[4]%
\>[4]{}[\mskip1.5mu (\Conid{ByteString},\Conid{ByteString})\mskip1.5mu]\to \Conid{Redis}\;(\Conid{R} \uplus \Conid{Status}){}\<[66]%
\>[66]{}~~,{}\<[E]%
\\
\>[B]{}\Varid{\Conid{Hedis}.hgetall}\mathbin{::}\Conid{ByteString}\to {}\<[E]%
\\
\>[B]{}\hsindent{4}{}\<[4]%
\>[4]{}\Conid{Redis}\;(\Conid{R} \uplus [\mskip1.5mu (\Conid{ByteString},\Conid{ByteString})\mskip1.5mu])~~.{}\<[E]%
\ColumnHook
\end{hscode}\resethooks
Proper implementations of them in \Edis{} should accept or return a
\emph{heterogeneous list}~\cite{hetero} --- a list whose elements can be of
different types. We also leave such functions as a future work.

\paragraph{Not All Safe Redis Programs Can Be Typechecked.}
Enforcing a typing discipline rules out some erroneous programs, and reduces the number of programs that are allowed. Like all type
systems, our type system takes a conservative estimation: there are bound to be
some \Redis{} programs that are not typable in our type system, but do not
actually throw a type error. We demand that elements in our lists must be
of the same type, for example, while a \Redis{} program could store in a list
different types of data, encoded as strings, and still work well.

One innate limitation is that we cannot allow dynamic generation of keys. In
\Hedis{}, the Haskell program is free to generate arbitrary sequence of keys
to be used in the data store, which is in general not possible due to the
static nature of \Edis{}.

\paragraph{Transactions.} Commands in \Redis{} can be wrapped in
\emph{transactions}. \Redis{} offers two promises regarding commands in a
transaction. Firstly, all commands in a transaction are serialized and
executed sequentially, without interruption from another client. Secondly,
either all of the commands or none are processed.

Support of transactions in \Edis{} is a future work.
We expect that there would not be too much difficulty --- in an early
experiment, we have implemented a runtime type checker specifically targeting
\Redis{} transactions, and we believe that the experience should be applicable
to static type checking as well.

%% file: sections/Conclusions.tex
\section{Conclusions and Related Work}
\label{sec:conclusions}

By exploiting various recent extensions and type-level programming techniques,
we have designed a domain-specific embedded language \Edis{} which enforces
typing disciplines that keep track of available keys and their types. The
type of a command clearly specifies which keys must or must not present in
the data store, what their types ought to be, as well as how the keys and types are updated after the execution. A program can be constructed only if it does
not throw a runtime type error when it is run with a store whose status matches
its precondition and when it is the sole client interacting with the store.
The type also serves as documentation of the
commands. We believe that it is a neat case study of application of type-level programming.

\Redis{} identifies itself as a data structure store/server, rather than a
database. Nevertheless, there has been attempts to design DSELs for relational
databases that guarantee all queries made are safe. Among them, {\sc HaskellDB}~\cite{haskelldb,haskelldbimproved} dynamically generates, from monad comprehensions, SQL queries to be executed on ODBC database servers.
With the expressiveness of dependent types, Oury and Swierstra~\cite{pi}
build a DSEL in Agda for relational algebra. Eisenberg and
Weirich~\cite{singletons} ported the result to Haskell using singleton types, after GHC introduced more features facilitating type-level programming.

None of the type-level programming techniques we used in this paper are new.
Indexed monads (also called parameterized monads) have been introduced by
Atkey~\cite{indexedmonad}. McBride~\cite{kleisli} showed how to construct
indexed free-monads from Hoare Logic specifications. Kiselyov et
al.~\cite{typefun} used indexed monads to track the locks held among a given
finite set. The same paper also demonstrated implementation of a variety of
features including memorization, generic maps, session types, typed
\texttt{printf} and \texttt{sprintf}, etc., by type-level programming.
Before the introduction of type families, Kiselyov and
Shan~\cite{staticresources} used type classes and functional dependencies to
implement type-level functions, and showed that they are sufficient to track
resources in device drivers.

Lindley and McBride~\cite{phasedistinction} provided a thorough analysis and summary of the dependent-type-like features currently in Haskell, and
compared them with dependently typed languages without phase distinction such as Agda. It turns out that GHC's constraint solver works surprisingly well as an
automatic theorem prover.